\documentclass[12pt]{iopart}

\usepackage{enumerate}
\usepackage{amsfonts}
\usepackage{amssymb}
\usepackage{amsthm}

\newcommand{\supp}{\mathrm{supp}}
\newcommand{\rank}{\mathrm{rank}}
\newcommand{\id}{\mathrm{Id}}

\newcommand{\shs}{\hspace{1pt}}

\newcounter{defin}  \newcounter{lemma}  \newcounter{theorem}
\newcounter{property} \newcounter{corol}  \newcounter{remark} \newcounter{example}

\newenvironment{lemma}{\par\refstepcounter{lemma}
     \textbf{Lemma \thelemma.} }{\rm\par}
\newenvironment{theorem}{\par\refstepcounter{theorem}
     \textbf{Theorem \thetheorem.}\ }{\rm\par}
\newenvironment{property}{\par\refstepcounter{property}
     \textbf{Proposition \theproperty.}\ }{\rm\par}
\newenvironment{corollary}{\par\refstepcounter{corol}
     \textbf{Corollary \thecorol.} }{\rm\par}

\newenvironment{remark}{\par\refstepcounter{remark}
     \textbf{Remark \theremark.}}{\rm\par}

\begin{document}

\title[Uniform continuity bounds for information characteristics of quantum channels]{Uniform continuity bounds for information characteristics of quantum channels depending on  input dimension and on input energy}

\author{M.E. Shirokov}

\address{Steklov Mathematical Institute, Moscow Institute of Physics and Technology, email:msh@mi.ras.ru}
\vspace{10pt}
\begin{indented}
\item[]October 2018
\end{indented}

\begin{abstract}
We obtain continuity bounds for basic  information characteristics of quantum channels depending on their input dimension (if it is finite) and on the input energy bound (if the input dimension is infinite). We pay a special attention to the case of a multi-mode quantum oscillator as an input system.

First, we prove  continuity bounds for the output conditional mutual information for a single channel and for $n$ copies of a channel.

Then we obtain  estimates for variation of the output Holevo quantity with respect to simultaneous variations of a channel and of an input ensemble.

As a result, tight and close-to-tight continuity bounds for basic capacities of quantum channels depending on the input dimension are obtained. They  complement the Leung-Smith continuity bounds depending on the output dimension.

Finally, we obtain tight and close-to-tight continuity bounds for basic capacities of infinite-dimensional energy-constrained channels with respect to the energy-constrained Bures distance generating the strong  convergence of quantum channels.
\end{abstract}

%
\noindent{\it Keywords}:  quantum channel capacities, quantum conditional mutual information, the Holevo quantity, ensemble of quantum states, energy-constrained Bures distance, strong convergence of quantum channels, multi-mode quantum oscillator
%

\submitto{\JPA}
%
%
%

\section{Introduction}

In study of informational properties of quantum channels it is necessary to take into account all possible perturbations of quantum states and quantum channels unavoidable in real physical scenarios. Mathematically, this means that we have to quantitatively analyse continuity of basic information characteristics (in particular, channel capacities) with respect to appropriate metrics on the sets of quantum states and quantum channels.

Leung and Smith obtained in \cite{L&S} (uniform) continuity bounds (estimates for variations) for basic capacities of  quantum channels depending of their output dimension.
The appearance of the output dimension in these and some other continuity bounds for information characteristics of a quantum channel is natural, since such characteristics are typically expressed via entropic quantities of output states of a channel (so,  application of  Fannes' type continuity bounds  gives the factor proportional to the logarithm of the output dimension  \cite{ A&F,Aud,Fannes, W-CB}).

At the same time, it is the input dimension of a channel that determines the range of its information capacities, while the formal output dimension may be substantially greater than the real dimension of a channel output. So, it is reasonable to conjecture that the input dimension also determines the rate of variations of capacities regardless of the output dimension.

Speaking about capacities of channels between infinite-dimensional quantum systems we have to impose energy constraints on states used for encoding information \cite{H-SCI,H-c-w-c,Wilde+}. In \cite{SCT,W-EBN} continuity bounds for classical and quantum capacities are obtained  for energy-constrained
infinite-dimensional channels with \emph{bounded energy amplification factor}. In fact, these continuity bounds depend on the output average energy of
states used for encoding information. Nevertheless, the range of information capacities is completely determined by the input  average energy of
code-states, which plays, roughly speaking, the role of input dimension \cite{W-CB}.  So, we may expect that the input energy bound also determines the rate of variations of capacities and of other  characteristics of infinite-dimensional channels  regardless of the output energy (which may be unbounded or equal to $+\infty$).

In this paper we confirm both these conjectures by deriving  continuity bounds for several information characteristics of quantum channels (in particular, for all basic capacities) depending on their input dimension (if it is finite) and on the  input energy bound (if the input dimension is infinite) provided that the Hamiltonian $H_A$ of the input system satisfies the condition
$$
\lim_{\lambda\rightarrow0^+}\left[\mathrm{Tr}\, e^{-\lambda H_A}\right]^{\lambda}=1,
$$
which holds, in particular, if the input system is a multi-mode quantum oscillator \cite{H-SCI,W&C}. This case plays a central role in quantum information theory. It is considered separately after formulations of general results.

We will consider the case of finite input dimension and the case of finite input energy simultaneously excepting the last Sections 7 and 8 devoted, respectively, to the first and to the second case.

In Section 3 we describe appropriate metrics on the set of quantum channels. In particular, we introduce the \emph{energy-constrained Bures distance} between quantum channels used in the "finite input energy" part of the paper.

Then we prove  special continuity bounds for the extented quantum conditional mutual information (Lemma \ref{S-CMI-CB} in Section 4), which are our main technical tools.

In Section 5  we obtain  continuity bounds for the output conditional mutual information $I(B\!:\!D|C)_{\Phi\otimes\id_{CD}(\rho)}$ with respect to simultaneous variations  of a channel  $\Phi:A\rightarrow B$ and of an input state  $\rho_{ACD}$.  We  also derive  continuity bound for the function $\Phi\mapsto I(B^n\!:\!D|C)_{\Phi^{\otimes n}\otimes\id_{CD}(\rho)}$ for any natural $n$ by using the Leung-Smith telescopic trick.

In Section 6  we analyse continuity properties of the output Holevo quantity $\chi(\Phi(\mu))$ -- the Holevo quantity of the ensemble $\Phi(\mu)$ obtained by action of a channel $\Phi$ on a given (discrete or continuous) ensemble $\mu$ of input states. We obtain  estimates for variation of $\chi(\Phi(\mu))$ with respect to simultaneous variations  of a channel $\Phi$ and of an input ensemble $\mu$.

In Section 7 the above results are applied to obtain tight and close-to-tight continuity bounds for basic capacities of quantum channels depending on their input dimension. They complement the Leung-Smith continuity bounds (depending on the output dimension).

In Section 8  we obtain tight and close-to-tight continuity bounds for basic capacities of infinite-dimensional energy-constrained channels with respect to the energy-constrained Bures distance generating the strong (pointwise) convergence of quantum channels. These continuity bounds are valid for arbitrary quantum channels (in contrast to the continuity bounds obtained in \cite{SCT,W-EBN}) and do not depend on channel characteristics at all.

\section{Preliminaries}

\subsection{Basic notations and auxiliary lemmas}

Let $\mathcal{H}$ be a finite-dimensional or separable infinite-dimensional Hilbert space. Denote by
$\mathfrak{B}(\mathcal{H})$ the algebra of all bounded operators on $\mathcal{H}$ with the operator norm $\|\cdot\|$ and
by $\mathfrak{T}( \mathcal{H})$ the Banach space of all trace-class
operators on $\mathcal{H}$  with the trace norm $\|\!\cdot\!\|_1$. Let
$\mathfrak{S}(\mathcal{H})$ be  the set of quantum states (positive operators
in $\mathfrak{T}(\mathcal{H})$ with unit trace) \cite{H-SCI,Wilde}.

The \emph{Bures distance} between quantum states $\rho$ and $\sigma$ is defined as
\begin{equation}\label{B-d-s}
  \beta(\rho,\sigma)=\sqrt{2\left(1-\sqrt{F(\rho,\sigma)}\right)},
\end{equation}
where $F(\rho,\sigma)=\|\sqrt{\rho}\sqrt{\sigma}\|^2_1$ is the \emph{fidelity} of $\rho$ and $\sigma$ \cite{H-SCI,Wilde}. The following relations  between the Bures distance and the  trace-norm distance hold
\begin{equation}\label{B-d-s-r}
\textstyle\frac{1}{2}\|\rho-\sigma\|_1\leq\beta(\rho,\sigma)\leq\sqrt{\|\rho-\sigma\|_1}.
\end{equation}

Denote by $I_{\mathcal{H}}$ and $\id_{\mathcal{H}}$ the unit operator on a Hilbert space
$\mathcal{H}$ and  the identity
transformation of the Banach space $\mathfrak{T}(\mathcal{H})$ correspondingly.

If quantum systems $A$ and $B$ are described by Hilbert spaces  $\mathcal{H}_A$ and $\mathcal{H}_B$ then the composite quantum system $AB$ is described by the Hilbert space $\mathcal{H}_{AB}\doteq\mathcal{H}_A\otimes\mathcal{H}_B$. For a state $\rho_{AB}$ in $\mathfrak{S}(\mathcal{H}_{AB})$ denote by $\rho_{A}$ and $\rho_{B}$ its marginal states $\mathrm{Tr}_B\rho_{AB}$ and $\mathrm{Tr}_A\rho_{AB}$ correspondingly (here and in what follows $\mathrm{Tr}_X$ denotes the partial trace $\mathrm{Tr}_{\mathcal{H}_X}$ over the space $\mathcal{H}_X$).

The \emph{von Neumann entropy} $H(\rho)=\mathrm{Tr}\eta(\rho)$ of a
state $\rho\in\mathfrak{S}(\mathcal{H})$, where $\eta(x)=-x\log x$ if $x>0$ and $\eta(0)=0$,
is a  nonnegative concave lower semicontinuous function on the set $\mathfrak{S}(\mathcal{H})$, it is continuous on $\mathfrak{S}(\mathcal{H})$
if and only if $\,\dim\mathcal{H}<+\infty$ \cite{H-SCI,L-2}.

The \emph{quantum relative entropy} for states $\rho$ and
$\sigma$ in $\mathfrak{S}(\mathcal{H})$ is defined as
$$
H(\rho\shs\|\shs\sigma)=\sum_i\langle
i|\,\rho\log\rho-\rho\log\sigma\,|i\rangle,
$$
where $\{|i\rangle\}$ is the orthonormal basis of
eigenvectors of the state $\rho$ and it is assumed that
$H(\rho\shs\|\shs\sigma)=+\infty$ if $\,\supp\shs\rho$ is not
contained in $\,\supp\shs\sigma$ \cite{H-SCI,L-2} (the support $\supp\shs\rho$ of a positive operator $\rho$ is the orthogonal complement to its kernel).

The \emph{quantum mutual information} of a state $\,\rho_{AB}\,$ of a
bipartite quantum system  is defined as
\begin{equation*}
I(A\!:\!B)_{\rho}=H(\rho_{AB}\shs\Vert\shs\rho_{A}\otimes
\rho_{B})=H(\rho_{A})+H(\rho_{B})-H(\rho_{AB}),
\end{equation*}
where the second formula is valid if $\,H(\rho_{AB})\,$ is finite \cite{L-mi}.

Properties of the quantum relative entropy imply that $\,\rho\mapsto
I(A\!:\!B)_{\rho}\,$ is a nonnegative lower semicontinuous function on the set
$\mathfrak{S}(\mathcal{H}_{AB})$. It is well known that
\begin{equation}\label{MI-UB}
I(A\!:\!B)_{\rho}\leq 2\min\left\{H(\rho_A),H(\rho_B)\right\}
\end{equation}
for arbitrary state $\rho_{AB}$ and that
\begin{equation}\label{MI-UB+}
I(A\!:\!B)_{\rho}\leq \min\left\{H(\rho_A),H(\rho_B)\right\}
\end{equation}
for any separable state $\rho_{AB}$ \cite{L-mi,MI-B}.

The \emph{quantum conditional mutual information (QCMI)} of a state $\rho_{ABC}$ of a
finite-dimensional tripartite quantum system is defined as
\begin{equation}\label{cmi-d}
    I(A\!:\!B|C)_{\rho}\doteq
    H(\rho_{AC})+H(\rho_{BC})-H(\rho_{ABC})-H(\rho_{C}).
\end{equation}
This quantity is essentially used  in quantum
information theory \cite{H-SCI,Wilde}.

To avoid possible uncertainty
$"\infty-\infty"$ in (\ref{cmi-d}) in infinite dimensions one should define the QCMI for any state
$\rho_{ABC}$ by one of the equivalent expressions
\begin{equation}\label{cmi-e+}
\hspace{-30pt}     I(A\!:\!B|C)_{\rho}=\sup_{P_A}\left[\shs I(A\!:\!BC)_{Q_A\rho
                    Q_A}-I(A\!:\!C)_{Q_A\rho Q_A}\shs\right],\quad Q_A=P_A\otimes I_{BC},
\end{equation}
\begin{equation}\label{cmi-e++}
\hspace{-30pt}     I(A\!:\!B|C)_{\rho}=\sup_{P_B}\left[\shs I(B\!:\!AC)_{Q_B\rho
                   Q_B}-I(B\!:\!C)_{Q_B\rho Q_B}\shs\right],\quad Q_B=P_B\otimes I_{AC},
\end{equation}
where the suprema are taken over all finite rank projectors
$P_A$ in $\mathfrak{B}(\mathcal{H}_A)$ and $P_B$ in $\mathfrak{B}(\mathcal{H}_B)$  \cite{CMI}. We assume here that $I(X\!:\!Y)_{Q_X\rho
Q_X}=c I(X\!:\!Y)_{c^{-1} Q_X\rho
Q_X}$, where  $c=\mathrm{Tr}\shs Q_X\rho_{ABC}$.

Formulae (\ref{cmi-e+}) and
(\ref{cmi-e++}) define the same  lower semicontinuous function on the set
$\mathfrak{S}(\mathcal{H}_{ABC})$ inheriting all basic properties of the finite-dimensional QCMI defined by (\ref{cmi-d}) \cite[Theorem 2]{CMI}. We will use  the following relation (chain rule)
\begin{equation}\label{chain}
I(X\!:\!YZ|C)_{\rho}=I(X\!:\!Y|C)_{\rho}+I(X\!:\!Z|YC)_{\rho}
\end{equation}
valid for any state $\rho$ in $\mathfrak{S}(\mathcal{H}_{XYZC})$ (with possible values $+\infty$ on both sides).
Relation (\ref{chain}) can be proved by noting that it holds if the systems $X,Y,Z$ and $C$ are finite-dimensional and by using Corollary 9 in \cite{CMI}.

We will use the upper bound
\begin{equation}\label{CMI-UB}
I(A\!:\!B|C)_{\rho}\leq 2\min\left\{H(\rho_A),H(\rho_B),H(\rho_{AC}),H(\rho_{BC})\right\}
\end{equation}
valid for any state $\rho_{ABC}$. It directly follows from upper bound (\ref{MI-UB}) and the expression
$I(X\!:\!Y|C)_{\rho}=I(X\!:\!YC)_{\rho}-I(X\!:\!C)_{\rho}$, $X,Y=A,B$, which is a partial case of (\ref{chain}).

The QCMI (defined in (\ref{cmi-e+}),(\ref{cmi-e++})) satisfies the following relation
\begin{equation}\label{F-c-b}
\hspace{-20pt}\begin{array}{cc}
\left|p
I(A\!:\!B|C)_{\rho}+(1-p)I(A\!:\!B|C)_{\sigma}-I(A\!:\!B|C)_{p\rho+(1-p)\sigma}\right|\leq h_2(p)
\end{array}
\end{equation}
valid for $p\in(0,1)$ and any states $\rho_{ABC}$, $\sigma_{ABC}$ with finite $I(A\!:\!B|C)_{\rho}$, $I(A\!:\!B|C)_{\sigma}$, where $h_2(p)=\eta(p)+\eta(1-p)$ is the binary entropy \cite{CHI}.

We will repeatedly use the following simple lemmas.
\begin{lemma}\label{sl} \emph{If $\;U$ and $\,V$ are isometries from $\mathcal{H}$ into $\mathcal{H}'$ then}
$$
\|U\rho U^*-V\rho V^*\|_1\leq 2\|(U-V)\rho\|_1\leq 2\|U-V\|\;\;\textit{for any}\;\;\rho\in\mathfrak{S}(\mathcal{H}).
$$
\end{lemma}

\begin{lemma}\label{GWL} \cite{W-CB} \emph{If $f$ is a concave nonnegative function on $[0,+\infty)$ then for any positive $x< y$ and any $z\geq0$ the following inequality holds}
$$
xf(z/x)\leq yf(z/y).
$$
\end{lemma}

\subsection{Set of quantum states with bounded energy}

\subsubsection{General case}

Let $H_A$ be a positive (semi-definite) operator on a Hilbert space $\mathcal{H}_A$ treated as a Hamiltonian of a quantum system $A$. Let $E\geq E_0\doteq\inf\limits_{\|\varphi\|=1}\langle\varphi|H_A|\varphi\rangle$. Then
$$
\mathfrak{C}_{H_A,E}=\{\rho\in\mathfrak{S}(\mathcal{H}_A)\,|\,\mathrm{Tr} H_A\rho\leq E\}
$$
is a closed convex subset of $\mathfrak{S}(\mathcal{H}_A)$ consisting of states with the mean energy not exceeding $E$.

It is well known that the von Neumann entropy is continuous on the set $\mathfrak{C}_{H_A,E}$ for any $E> E_0$ if (and only if) the Hamiltonian  $H_A$ satisfies  the condition
\begin{equation}\label{H-cond}
  \mathrm{Tr}\, e^{-\lambda H_{A}}<+\infty\quad\textrm{for all}\;\lambda>0
\end{equation}
and that the maximal value of the entropy on this set is achieved at the \emph{Gibbs state} $\gamma_A(E)\doteq e^{-\lambda(E) H_A}/\mathrm{Tr} e^{-\lambda(E) H_A}$, where the parameter $\lambda(E)$ is determined by the equality $\mathrm{Tr} H_A e^{-\lambda(E) H_A}=E\mathrm{Tr} e^{-\lambda(E) H_A}$ \cite{W}. Condition (\ref{H-cond}) implies that $H_A$ is an unbounded operator having  discrete spectrum of finite multiplicity \cite{H-c-w-c}.

We will use the function $F_{H_A}(E)\doteq\sup_{\rho\in\mathfrak{C}_{H_{\!A},E}}H(\rho)=H(\gamma_A(E))$.
It is easy to show that $F_{H_A}$ is a strictly increasing concave function on $[E_0,+\infty)$ such that $F_{H_A}(E_0)=\log d_0$, where $d_0$ is the multiplicity of  $E_0$ \cite{W-CB,CHI}.

Let
\begin{equation}\label{F-bar}
 \bar{F}_{H_A}(E)=F_{H_A}(E+E_0)=H(\gamma_A(E+E_0))
\end{equation}
 and $\bar{F}^{-1}_{H_A}$ be the inverse function to the function $\bar{F}_{H_A}$. The above-stated properties of $F_{H_A}$ shows that
$\bar{F}^{-1}_{H_A}$ is an increasing function on $[\log d_0, +\infty)$ taking values in $[0, +\infty)$.\smallskip

\begin{lemma}\label{cl} \emph{Let  $\,\bar{E}\doteq E-E_0\geq 0$, $\,d\geq d_0\,$ and  $\;\gamma(d)\doteq \bar{F}_{H_A}^{-1}(\log d)$. Let $B$ be any system. If $\,\bar{E}\leq \gamma(d)$ then for any pure state $\rho$ in $\,\mathfrak{S}(\mathcal{H}_{AB})$ such that $\mathrm{Tr} H_A\rho_A\leq E$ there is a pure state $\sigma$ in $\,\mathfrak{S}(\mathcal{H}_{AB})$ such that
$\,\rank\shs \sigma_A\leq d$, $\,\mathrm{Tr} H_A\sigma_A\leq E$, $\,\textstyle\frac{1}{2}\|\rho-\sigma\|_1\leq \displaystyle\sqrt{\bar{E}/\gamma(d)}\,$ and
$$
\|\rho-\sigma\|_1\mathrm{Tr}\bar{H}_A\left[[\rho-\sigma]_{-}\right]_A\leq 2\bar{E},\quad \|\rho-\sigma\|_1\mathrm{Tr}\bar{H}_A\left[[\rho-\sigma]_{+}\right]_A\leq 2\bar{E},
$$
where $\bar{H}_A=H_A-E_0I_{A}$, $[\rho-\sigma]_-$ and $\,[\rho-\sigma]_+$  are, respectively, the negative and positive parts of the Hermitian operator $\,\rho-\sigma$.}
\end{lemma}\smallskip

\emph{Proof.}   Note first that
\begin{equation}\label{t-lb}
C\doteq\max_{1\leq k\leq d}\{\langle k|\bar{H}_A|k\rangle \}\geq \gamma(d)
\end{equation}
for any orthonormal set $\{|k\rangle\}_{k=1}^d\subset\mathcal{H}_A$.  Indeed, if $P_d=\sum_{k=1}^d |k\rangle\langle k|$ then
$\mathrm{Tr} P_d \bar{H}_A\leq Cd$. So, the entropy $\log d$ of the state $d^{-1}P_d$ does not exceed $F_{\bar{H}_A}(C)=\bar{F}_{H_A}(C)$.

Let $\rho=|\varphi\rangle\langle\varphi|$, where  $|\varphi\rangle=\sum_{k=1}^{+\infty} \sqrt{p_k}|\alpha_k\otimes \beta_k\rangle$  for some base $\{|\alpha_k\rangle\}$ and $\{|\beta_k\rangle\}$ in the Hilbert spaces
$\mathcal{H}_A$ and $\mathcal{H}_B$. We may assume that the basis
$\{|\alpha_k\rangle\}$ is reordered in such a way that the sequence $\{\langle \alpha_k|\bar{H}_A|\alpha_k\rangle\}_k$ is non-decreasing.

Let $|\psi\rangle=(1-\delta_d)^{-1/2}\sum_{k=1}^{d} \sqrt{p_k}|\alpha_k\otimes \beta_k\rangle$,  where $\delta_d=\sum_{k>d} p_k$, and $\sigma=|\psi\rangle\langle\psi|$.

Since $\mathrm{Tr} \bar{H}_A\rho_A\leq \bar{E}$, we have
$\delta_d \langle \alpha_d|\bar{H}_A|\alpha_d\rangle\leq \sum_{k>d} p_k \langle \alpha_k|\bar{H}_A|\alpha_k\rangle\leq \bar{E}$ and
hence
\begin{equation*}
    \delta_d\leq \bar{E}/\langle \alpha_d|\bar{H}_A|\alpha_d\rangle\leq \bar{E}/\gamma(d)\leq 1,
\end{equation*}
where the second inequality follows from (\ref{t-lb}). It is easy to see that the above inequality implies $\mathrm{Tr} \bar{H}_A\sigma_A\leq \bar{E}$, i.e. $\mathrm{Tr}H_A\sigma_A\leq E$.

Since $\langle\varphi|\psi\rangle=(1-\delta_d)^{1/2}$, we have $\,\textstyle\frac{1}{2}\|\rho-\sigma\|_1=\sqrt{1-|\langle\varphi|\psi\rangle|^2}=\sqrt{\delta_d}$.

Assume that $\rho\neq\sigma$ (otherwise all the assertions are trivial). By using spectral decomposition of the operator $\;\rho-\sigma=|\varphi\rangle\langle\varphi|-|\psi\rangle\langle\psi|$ one can show that
$[\rho-\sigma]_{\pm}$ are 1-rank operators corresponding to the  vectors
$|\gamma_{\pm}\rangle=p_{\pm}|\varphi\rangle+q_{\pm}|\psi\rangle$, where $p_{\pm}=\sqrt{(1\pm x)/(2x)}$ and
$q_{\pm}=-\sqrt{(1\mp x)/2x},\; x=\sqrt{\delta_d}$ (see  the proof of Theorem 1 in \cite{AFM}).
So, we have
$$
\begin{array}{c}
\mathrm{Tr}\bar{H}_A[[\rho-\sigma]_{\pm}]_A=\langle\gamma_{\pm}|\bar{H}_A\otimes I_B|\gamma_{\pm}\rangle=p_{\pm}^2\langle\varphi|\bar{H}_A\otimes I_B|\varphi\rangle+q_{\pm}^2\langle\psi|\bar{H}_A\otimes I_B|\psi\rangle
\\\\+2p_{\pm}q_{\pm}\Re\langle\varphi|\bar{H}_A\otimes I_B|\psi\rangle=\left[p_{\pm}^2+ 2p_{\pm}q_{\pm}(1-\delta_d)^{-1/2}+ q_{\pm}^2(1-\delta_d)^{-1}\right]S+p_{\pm}^2R\\\\
=\displaystyle \frac{1}{2x}\left[\frac{x^2S}{1\pm x}+(1\pm x)R\shs\right] \leq \displaystyle \frac{1}{2x}\left[x^2(1\mp x)\bar{E}+(1\pm x)\bar{E}\right]\leq \frac{\bar{E}}{x}=\frac{2\bar{E}}{\|\rho-\sigma\|_1},
\end{array}
$$
where $S=\sum_{k\leq d} p_k \langle \alpha_k|\bar{H}_A|\alpha_k\rangle$ and $R=\sum_{k> d} p_k \langle \alpha_k|\bar{H}_A|\alpha_k\rangle$.
In the last line we used the obvious inequalities $S\leq (1-x^2)\bar{E}$, $R\leq\bar{E}$ and $x\leq 1$. $\square$

In this paper we  essentially use the  modification of the Alicki-Fannes-Winter method (cf.\cite{A&F,W-CB,M&H}) adapted for the set of states with bounded energy  \cite{AFM}. This modification makes it possible to prove uniform continuity of any locally almost affine function\footnote{This means that  $\,|f(p\rho+(1-p)\sigma)-p f(\rho)-(1-p)f(\sigma)|\leq r(p)=o(1)\,$ as $\,p\rightarrow0^{+}$ for any $\rho$ and $\sigma$.} $\,f\,$ on the set
$$
\mathfrak{C}^{\,\mathrm{ext}}_{H_A,E}\doteq\{\shs\rho\in\mathfrak{S}(\mathcal{H}_{AB})\shs|\,\rho_A\in \mathfrak{C}_{H_A,E}\shs\}\qquad (B\; \textrm{ is any given system})
$$
such that $|f(\rho_{AB})|\leq C H(\rho_A)$ for some $C\in\mathbb{R}_+ $ provided that
\begin{equation}\label{H-cond++}
  F_{H_A}(E)=o\shs(\sqrt{E})\quad\textrm{as}\quad E\rightarrow+\infty.
\end{equation}
By Lemma 1 in \cite{AFM} condition (\ref{H-cond++}) holds if and only if
\begin{equation}\label{H-cond+}
  \lim_{\lambda\rightarrow0^+}\left[\mathrm{Tr}\, e^{-\lambda H_A}\right]^{\lambda}=1.
\end{equation}
Condition (\ref{H-cond+}) is  stronger than condition (\ref{H-cond}) (equivalent to  $\,F_{H_A}(E)=o\shs(E)\,$ as $\,E\rightarrow+\infty$) but the difference between these conditions  is not too large. In terms of the sequence $\{E_k\}$ of eigenvalues of $H_A$
condition (\ref{H-cond}) means that $\lim_{k\rightarrow\infty}E_k/\log k=+\infty$, while (\ref{H-cond+}) is valid  if $\;\liminf_{k\rightarrow\infty} E_k/\log^q k>0\,$ for some $\,q>2$ \cite[Proposition 1]{AFM}.

It is essential that condition (\ref{H-cond+})  holds for the Hamiltonian of a system of quantum oscillators considered in the next subsection.

\subsubsection{The $\ell$-mode quantum oscillator}

Let $A$ be the $\,\ell$-mode quantum oscillator with the frequencies $\,\omega_1,...,\omega_{\ell}\,$. In this case
$$
F_{H_A}(E)=\max_{\{x_i\}}\,\sum_{i=1}^{\ell}g(x_i/\hbar\omega_i-1/2),\quad E\geq E_0\doteq\frac{1}{2}\sum_{i=1}^{\ell}\hbar\omega_i,
$$
where $\,g(x)=(x+1)\log(x+1)-x\log x\,$ and the maximum is over all $\ell\textup{-}$tuples $x_1$,...,$x_{\ell}$  such that  $\sum_{i=1}^{\ell}x_i=E$ and $x_i\geq\frac{1}{2}\hbar\omega_i$ \cite[Ch.12]{H-SCI}. It is shown in \cite{CHI} that
\begin{equation}\label{F-ub}
F_{H_A}(E)\leq F_{\ell,\omega}(E)\doteq \ell\log \frac{E+E_0}{\ell E_*}+\ell,\quad E_*=\left[\prod_{i=1}^{\ell}\hbar\omega_i\right]^{1/\ell}\!\!,\vspace{-5pt}
\end{equation}
(throughout the paper $\log$ is the natural logarithm) and that upper bound (\ref{F-ub}) is $\varepsilon$-sharp for large $E$. So, the function
\begin{equation}\label{F-ub+}
\bar{F}_{\ell,\omega}(E)\doteq F_{\ell,\omega}(E+E_0)=\ell\log \frac{E+2E_0}{\ell E_*}+\ell,\vspace{-5pt}
\end{equation}
is an $\varepsilon$-sharp upper bound on the function $\bar{F}_{H_A}(E)\doteq F_{H_A}(E+E_0)$. By employing the function $\bar{F}_{\ell,\omega}$ one can  define
the sequence of positive numbers
\begin{equation}\label{gamma-h}
\hat{\gamma}(d)\doteq\bar{F}^{-1}_{\ell,\omega}(\log d)=(\ell/e)E_*\sqrt[\ell]{d}-2E_0,\quad d> e^{\bar{F}_{\ell,\omega}(0)},
\end{equation}
which can be used instead of the sequence $\gamma(d)$ in Lemma \ref{cl}. It is clear that $\,\hat{\gamma}(d)\leq\gamma(d)\,$ for all $d$.

\subsection{Some facts about ensembles of quantum states}

\subsubsection{Discrete ensembles}

A finite or
countable set $\{\rho_{i}\}$ of quantum states
and a corresponding probability distribution $\{p_{i}\}$ is usually called a
\textit{discrete ensemble} and denoted by $\{p_{i},\rho_{i}\}$. The state
$\bar{\rho}\doteq\sum_{i}p_{i}\rho_{i}$ is called the \emph{average state} of this  ensemble \cite{H-SCI,Wilde}.

The \emph{Holevo quantity} of an ensemble $\{p_i,\rho_i\}_{i=1}^m$ of $\,m\leq+\infty$ quantum states is defined as
$$
\chi\left(\{p_i,\rho_i\}_{i=1}^m\right)\doteq \sum_{i=1}^m p_i H(\rho_i\|\bar{\rho})=H(\bar{\rho})-\sum_{i=1}^m p_i H(\rho_i),
$$
where the second expression is valid if $H(\bar{\rho})<+\infty$. This quantity is an upper bound for the classical information obtained from the ensemble by quantum measurements \cite{H-73}. It is essentially used in analysis of information properties of quantum channels and systems \cite{H-SCI,Wilde}.

Assume that  $\mathcal{H}_A=\mathcal{H}$ and that $\,\{|i\rangle\}_{i=1}^m$ is a basis in a Hilbert space $\mathcal{H}_B$. Then 
\begin{equation}\label{chi-rep}
\chi(\{p_i,\rho_i\}_{i=1}^m)=I(A\!:\!B)_{\hat{\rho}},\textrm{ where }\,\hat{\rho}_{AB}=\sum_{i=1}^m p_i\rho_i\otimes |i\rangle\langle i|.
\end{equation}
The state $\hat{\rho}_{AB}$  is called the \emph{$qc$-state} determined by the ensemble $\{p_i,\rho_i\}_{i=1}^m$ \cite{Wilde}.

In analysis of continuity of the Holevo quantity we will use three  measures of divergence between ensembles $\mu=\{p_i,\rho_i\}$ and $\nu=\{q_i,\sigma_i\}$ described in detail in \cite{O&C,CHI}.

The quantity
\begin{equation*}
D_0(\mu,\nu)\doteq\frac{1}{2}\sum_i\|\shs p_i\rho_i-q_i\sigma_i\|_1
\end{equation*}
is a true metric on the set of all  ensembles of quantum states considered as \emph{ordered} collections of states with the corresponding probability distributions.

The main advantage of $D_0$ is a direct computability, but from the quantum information point of view it is natural to consider an ensemble of quantum states $\{p_i,\rho_i\}$ as a discrete probability measure $\sum_i p_i\delta(\rho_i)$  on the set $\mathfrak{S}(\mathcal{H})$ (where $\delta(\rho)$ is the Dirac measure concentrated at a state $\rho$) rather than ordered (or disordered) collection of states. If we want to identify ensembles corresponding to the same probability measure then we have to use the factorization of $D_0$, i.e. the quantity
 \begin{equation}\label{f-metric}
D_*(\mu,\nu)\doteq \inf_{\mu'\in \mathcal{E}(\mu),\nu'\in \mathcal{E}(\nu)}D_0(\mu',\nu')
\end{equation}
as a measure of divergence between ensembles $\mu=\{p_i,\rho_i\}$ and $\nu=\{q_i,\sigma_i\}$, where $\mathcal{E}(\mu)$ and $\mathcal{E}(\nu)$ are the sets
of all countable ensembles corresponding to the measures $\sum_i p_i\delta(\rho_i)$ and $\sum_i q_i\delta(\sigma_i)$ respectively.

It is shown in \cite{CHI} that the factor-metric $D_*$ coincides with the EHS-distance $D_{\mathrm{ehs}}$ between ensembles of quantum states proposed by Oreshkov and Calsamiglia in \cite{O&C} and that $D_*$ induces the topology of weak convergence on the set of all ensembles (interpreted as probability measures).\footnote{This means that a  sequence $\{\{p^n_i,\rho^n_i\}\}_n$  converges to an ensemble $\{p^0_i,\rho^0_i\}$ w.r.t. the metric $\,D_*\,$ if and only if
$\,\lim_{n\rightarrow\infty}\sum_i p^n_if(\rho^n_i)=\sum_i p^0_if(\rho^0_i)\,$
for any continuous bounded function $f$ on $\,\mathfrak{S}(\mathcal{H})$.}  It is obvious that
\begin{equation}\label{d-ineq}
D_*(\mu,\nu)\leq D_0(\mu,\nu)
\end{equation}
for any ensembles $\mu$ and $\nu$.

We will also use the Kantorovich distance
\begin{equation}\label{K-D-d}
D_K(\mu,\nu)=\frac{1}{2}\inf_{\{P_{ij}\}}\sum P_{ij}\|\rho_i-\sigma_j\|_1
\end{equation}
between ensembles $\mu=\{p_i,\rho_i\}$ and $\nu=\{q_i,\sigma_i\}$ of quantum states, where the infimum is taken over all joint probability distributions $\{P_{ij}\}$  such that $\sum_jP_{ij}=p_i$ for all $i$ and $\sum_iP_{ij}=q_j$ for all $j$. Since $D_*=D_{\mathrm{ehs}}$, it follows from the result in \cite{O&C} that
\begin{equation}\label{d-ineq+}
  D_*(\mu,\nu)\leq D_K(\mu,\nu)
\end{equation}
for any discrete ensembles $\mu$ and $\nu$.

The Kantorovich distance has a natural extension to the set of all generalized (continuous) ensembles. It generates the topology of weak convergence on this set (see the next subsection).

If $\,\mu$  and  $\,\nu$ are discrete ensembles of states in $\mathfrak{S}(\mathcal{H})$, where $\,d\doteq\dim\mathcal{H}<+\infty$, then Proposition 16 in \cite{CHI} implies that
\begin{equation}\label{CHI-CB+}
\left|\chi(\mu)-\chi(\nu)\right|\leq \varepsilon
\log d+2g(\varepsilon),
\end{equation}
where $\;\varepsilon=D_*(\mu,\nu)\,$ and
$\,g(\varepsilon)=(1+\varepsilon)h_2\!\left(\frac{\varepsilon}{1+\varepsilon}\right)$. Since $g(\varepsilon)$ is an increasing function, it follows from (\ref{d-ineq}) and (\ref{d-ineq+}) that inequality (\ref{CHI-CB+}) remains valid for $\,\varepsilon=D_0(\mu,\nu)\,$ and for $\,\varepsilon=D_K(\mu,\nu)$. Continuity bound (\ref{CHI-CB+}) with $\,\varepsilon=D_K(\mu,\nu)\,$ is a refined version of the continuity bound for the Holevo quantity obtained  in \cite{O&C}.

\subsubsection{Generalized (continuous) ensembles}

In the study of infinite-dimensional quantum systems the notion of \textit{generalized (continuous) ensemble} defined as
a Borel probability measure on the set of quantum states is widely used \cite{H-SCI,H-Sh-2}. We denote by $\mathcal{P}(\mathcal{H})$ the set of all Borel probability measures on $\mathfrak{S}(\mathcal{H})$ equipped with the topology of weak convergence
\cite{Bog,Par}.\footnote{A sequence $\{\mu_n\}$ of measures weakly converges to a measure $\mu_0$ if
$\,\lim_{n\rightarrow\infty}\int f(\rho)\mu_n(d\rho)=\int f(\rho)\mu_0(d\rho)\,$
for any continuous bounded function $f$ on $\,\mathfrak{S}(\mathcal{H})$.}
 The set $\mathcal{P}(\mathcal{H})$ can be treated as a complete
separable metric space \cite{Par}. It contains the dense subset $\mathcal{P}_0(\mathcal{H})$ of discrete measures (corresponding to discrete ensembles). The average state of a generalized
ensemble $\mu \in \mathcal{P}(\mathcal{H})$ is defined as the barycenter of the measure
$\mu $, that is
$\bar{\rho}(\mu)=\int_{\mathfrak{S}(\mathcal{H})}\rho \mu (d\rho )$.

For an ensemble $\mu \in \mathcal{P}(\mathcal{H}_{A})$ its image $\Phi(\mu) $
under a quantum channel $\Phi :A\rightarrow B\,$ is defined as the
ensemble in $\mathcal{P}(\mathcal{H}_{B})$ corresponding to the measure $\mu
\circ \Phi ^{-1}$ on $\mathfrak{S}(\mathcal{H}_{B})$, i.e. $\,\Phi (\mu )[%
\mathfrak{S}_{B}]=\mu[\Phi ^{-1}(\mathfrak{S}_{B})]\,$ for any Borel subset $%
\mathfrak{S}_{B}\subseteq \mathfrak{S}(\mathcal{H}_{B})$, where $\Phi ^{-1}(%
\mathfrak{S}_{B})$ is the pre-image of $\mathfrak{S}_{B}$ under the map $%
\Phi $. If $\mu =\{p _{i},\rho _{i}\}$ then  $\Phi (\mu)=\{p _{i},\Phi(\rho_{i})\}$.

The Holevo quantity of a
generalized ensemble $\mu \in \mathcal{P}(\mathcal{H})$ is defined as
\begin{equation*}
\chi(\mu)=\int_{\mathfrak{S}(\mathcal{H})} H(\rho\shs \|\shs \bar{\rho}(\mu))\mu (d\rho )=H(\bar{\rho}(\mu
))-\int_{\mathfrak{S}(\mathcal{H})} H(\rho)\mu (d\rho ),  
\end{equation*}%
where the second formula is valid under the condition $H(\bar{\rho}(\mu))<+\infty$ \cite{H-Sh-2}.

The Kantorovich distance (\ref{K-D-d}) is extended to generalized ensembles $\mu$ and $\nu$ by the expression
\begin{equation}\label{K-D-c}
D_K(\mu,\nu)=\frac{1}{2}\inf_{\Lambda\in\Pi(\mu,\nu)}\int_{\mathfrak{S}(\mathcal{H})\times\mathfrak{S}(\mathcal{H})}\|\rho-\sigma\|_1\Lambda(d\rho,d\sigma),
\end{equation}
where $\Pi(\mu,\nu)$ is the set of all Borel probability measures on $\mathfrak{S}(\mathcal{H})\times\mathfrak{S}(\mathcal{H})$ with the marginals $\mu$ and $\nu$. By noting that  $\frac{1}{2}\|\rho-\sigma\|_1\leq 1$ for any states $\rho$ and $\sigma$, we see that the Kantorovich distance (\ref{K-D-c}) generates the weak convergence on the set $\mathcal{P}(\mathcal{H})$  \cite{Bog}.

\section{Bures distance and Energy-Constrained Bures distance between quantum channels}

A \emph{quantum channel} $\,\Phi$ from a quantum system $A$ to a quantum system
$B$ is a linear completely positive trace preserving map from
$\mathfrak{T}(\mathcal{H}_A)$ into $\mathfrak{T}(\mathcal{H}_B)$ \cite{H-SCI,Wilde}.

For any  quantum channel $\,\Phi:A\rightarrow B\,$ the Stinespring theorem implies existence of a Hilbert space
$\mathcal{H}_E$ and  an isometry
$V_{\Phi}:\mathcal{H}_A\rightarrow\mathcal{H}_B\otimes\mathcal{H}_E$ such
that
\begin{equation}\label{St-rep}
\Phi(\rho)=\mathrm{Tr}_{E}V_{\Phi}\rho V_{\Phi}^{*},\quad
\rho\in\mathfrak{T}(\mathcal{H}_A).
\end{equation}
The quantum  channel
\begin{equation}\label{c-channel}
\mathfrak{T}(\mathcal{H}_A)\ni
\rho\mapsto\widehat{\Phi}(\rho)=\mathrm{Tr}_{B}V_{\Phi}\rho
V_{\Phi}^{*}\in\mathfrak{T}(\mathcal{H}_E)
\end{equation}
is called \emph{complementary} to the channel $\Phi$
\cite[Ch.6]{H-SCI}.

The set of quantum channels with finite input dimension $\dim\mathcal{H}_A$ is typically equipped with the metric induced by
the \emph{diamond norm}
\begin{equation}\label{d-norm}
\|\Phi\|_{\diamond}\doteq \sup_{\rho\in\mathfrak{S}(\mathcal{H}_{AR})}\|\Phi\otimes \id_R(\rho)\|_1\vspace{-5pt}
\end{equation}
of a Hermitian-preserving superoperator $\Phi:\mathfrak{T}(\mathcal{H}_A)\rightarrow\mathfrak{T}(\mathcal{H}_B)$, where $R$ is any system \cite{Kit},\cite[Ch.9]{Wilde}. This norm  coincides with the norm of complete boundedness of the dual map $\Phi^*:\mathfrak{B}(\mathcal{H}_B)\rightarrow\mathfrak{B}(\mathcal{H}_A)$ to the map $\Phi$ \cite{Paul}.

For our purposes it is more convenient to use the equivalent metric on the set of quantum channels called the \emph{Bures distance}. It is defined for given channels $\Phi$  and $\Psi$ from $A$ to $B$ as
\begin{equation}\label{b-dist+}
\beta(\Phi,\Psi)=\sup_{\rho\in\mathfrak{S}(\mathcal{H}_{AR})} \beta\!\left(\Phi\otimes \id_R(\rho),\Psi\otimes \id_R(\rho)\right),
\end{equation}
where $\beta(\cdot,\cdot)$ in the r.h.s. is the Bures distance between quantum states defined in (\ref{B-d-s}) and $R$ is any system. This metric is related to the notion of \emph{operational fidelity} for quantum channels introduced in \cite{B&Co}. It is studied in detail in \cite{Kr&W+,Kr&W}. In particular, it is shown in \cite{Kr&W} that the Bures distance (\ref{b-dist+}) can be also defined as
\begin{equation}\label{b-dist}
\beta(\Phi,\Psi)=\inf\|V_{\Phi}-V_{\Psi}\|,
\end{equation}
where the infimum is over all common Stinespring representations
\begin{equation}\label{c-S-r}
\Phi(\rho)=\mathrm{Tr}_E V_{\Phi}\rho V^*_{\Phi}\quad\textrm{and}\quad\Psi(\rho)=\mathrm{Tr}_E V_{\Psi}\rho V^*_{\Psi},
\end{equation}
and that the infimum in (\ref{b-dist}) is attainable. Definition (\ref{b-dist+}) and the relations (\ref{B-d-s-r}) imply that
\begin{equation}\label{DB-rel}
\textstyle\frac{1}{2}\|\Phi-\Psi\|_{\diamond}\leq\beta(\Phi,\Psi)\leq\sqrt{\|\Phi-\Psi\|_{\diamond}}.
\end{equation}
This shows the equivalence of the Bures distance and the diamond-norm distance on the set of all channels between given quantum systems.

In the case $\dim\mathcal{H}_A=+\infty$ the diamond-norm distance (the Bures distance) becomes singular: there are infinite-dimensional quantum channels with arbitrarily close physical parameters  having the diamond-norm distance equal to $2$ \cite{W-EBN}. In this case it is natural to use the distance between quantum channels induced by the \emph{energy-constrained diamond norm}
\begin{equation}\label{ecd}
\|\Phi\|^E_{\diamond}\doteq \sup_{\rho\in\mathfrak{S}(\mathcal{H}_{AR}), \mathrm{Tr} H_A\rho_A\leq E}\|\Phi\otimes \id_R(\rho)\|_1,\quad E>E_0,
\end{equation}
of a Hermitian-preserving superoperator $\Phi:\mathfrak{T}(\mathcal{H}_A)\rightarrow\mathfrak{T}(\mathcal{H}_B)$, where $H_A$ is the Hamiltonian of the input system $A$ and $E_0\doteq\inf\limits_{\|\varphi\|=1}\langle\varphi|H_A|\varphi\rangle$ \cite{SCT,W-EBN} (slightly different energy-constrained diamond norms are used in \cite{Pir}). If $H_A$ is an unbounded operator having discrete spectrum of finite multiplicity (in particular, if it satisfies condition (\ref{H-cond})) then  any of the norms (\ref{ecd}) generates the strong (pointwise) convergence of quantum channels, i.e.
vanishing of $\,\|\Phi_n-\Phi_0\|_{\diamond}^E\,$ as $\,n\rightarrow+\infty$
is equivalent to
\begin{equation}\label{sc-def}
\lim_{n\rightarrow+\infty}\Phi_n(\rho)=\Phi_0(\rho)\quad \forall\rho\in\mathfrak{S}(\mathcal{H}_A)
\end{equation}
for a sequence $\{\Phi_n\}$ of quantum channels from $A$ to any system $B$ \cite[Proposition 3]{SCT}. Note that the strong convergence is substantially weaker than the diamond norm convergence if the systems $A$ and $B$ are infinite-dimensional.

In this paper we will use the \emph{energy-constrained Bures distance}
\begin{equation}\label{ec-b-dist}
\hspace{-20pt}\beta_E(\Phi,\Psi)=\sup_{\rho\in\mathfrak{S}(\mathcal{H}_{AR}), \mathrm{Tr} H_A\rho_A\leq E} \beta(\Phi\otimes \id_R(\rho),\Psi\otimes \id_R(\rho)), \quad E> E_0,
\end{equation}
between quantum channels $\Phi$  and $\Psi$ from $A$ to $B$, where $R$ is any system.

Definitions (\ref{ecd}),(\ref{ec-b-dist}) and the relations (\ref{B-d-s-r}) imply that
\begin{equation}\label{DB-rel+}
\textstyle\frac{1}{2}\|\Phi-\Psi\|^E_{\diamond}\leq\beta_E(\Phi,\Psi)\leq\sqrt{\|\Phi-\Psi\|^E_{\diamond}}.
\end{equation}
The properties of the energy-constrained Bures distance used below are collected in the following

\begin{property}\label{ec-b-d-p}
\emph{Let $H_A$ be any positive densely defined operator on $\mathcal{H}_A$. For any $E> E_0$ the function $(\Phi,\Psi)\mapsto\beta_E(\Phi,\Psi)$ defined in (\ref{ec-b-dist}) is a real metric on the set of all quantum channels from $A$ to $B$ which can be represented as follows
\begin{equation}\label{ec-b-dist+}
\beta_E(\Phi,\Psi)=\inf \sup_{\rho\in\mathfrak{S}(\mathcal{H}_{A}),\mathrm{Tr} H_A\rho\leq E}\sqrt{\mathrm{Tr}(V_{\Phi}-V_{\Psi})\rho\shs(V^*_{\Phi}-V^*_{\Psi})},
\end{equation}
where the infimum is over all common Stinespring representations (\ref{c-S-r}). For any given channels $\,\Phi$ and $\,\Psi$ the following properties hold:}
\begin{enumerate}[a)]
  \item \emph{$\beta_E(\Phi,\Psi)$ tends to $\beta(\Phi,\Psi)$ as $E\rightarrow+\infty$;}
  \item \emph{the function $E\mapsto\beta_E(\Phi,\Psi)$ is concave and nondecreasing on $\,[E_0,+\infty)$;}
  \item \emph{for any $E\geq E_0$ the infimum in (\ref{ec-b-dist+}) is attained at a pair $(V_{\Phi},V_{\Psi})$ of isometries forming representation (\ref{c-S-r}).}
\end{enumerate}

\emph{If the operator $H_A$ satisfies condition (\ref{H-cond}) then for any $E>E_0$ the metric $\beta_E$ generates the strong (pointwise) convergence of quantum channels, i.e. vanishing of $\beta_E(\Phi_n,\Phi_0)$ as $\,n\rightarrow+\infty$ for a sequence $\{\Phi_n\}$ of quantum channels from $A$ to any system $B$
is equivalent to (\ref{sc-def}).}
\end{property}\smallskip

\emph{Proof.} The coincidence of (\ref{ec-b-dist}) and (\ref{ec-b-dist+}) along with the property c) follow from the proof of Theorem 1 in \cite{Kr&W}. For readers convenience we
present a simplified version of these arguments  for the case $\mathcal{A}=\mathfrak{B}(\mathcal{H})$ in the Appendix.

Property a) and the monotonicity of $\beta_E$ are obvious. To prove the concavity take any $E_1,E_2>E_0$. For arbitrary $\varepsilon>0$ there exist states $\rho_1$ and $\rho_2$ in $\mathfrak{S}(\mathcal{H}_{AR})$ such that
$$
\mathrm{Tr} H_A[\rho_i]_A\leq E_i\;\;\textrm{and} \;\;\beta^2_{E_i}(\Phi,\Psi)\leq \beta^2(\Phi\otimes \id_R(\rho_i),\Psi\otimes \id_R(\rho_i))+\varepsilon,\;\;i=1,2.
$$
By invariance of the Bures distance (\ref{B-d-s}) with respect to unitary transformation of both states we may assume that $\supp \mathrm{Tr}_A \rho_1\,\bot\,\supp \mathrm{Tr}_A \rho_2$ and hence\break $\supp \,\Theta_1\otimes \id_R(\rho_1)\,\bot\,\supp\, \Theta_2\otimes \id_R(\rho_2)$ for any channels $\Theta_1$ and $\Theta_2$. Thus, 
$$
\begin{array}{rl}
\beta^2_{\bar{E}}(\Phi,\Psi)\!\!\!&\geq
\beta^2(\Phi\otimes \id_R(\bar{\rho}),\Psi\otimes \id_R(\bar{\rho}))=\frac{1}{2}\beta^2(\Phi\otimes \id_R(\rho_1),\Psi\otimes \id_R(\rho_1))\\\\&+\,\frac{1}{2}\beta^2(\Phi\otimes \id_R(\rho_2),\Psi\otimes \id_R(\rho_2))\geq\frac{1}{2}(\beta^2_{E_1}(\Phi,\Psi)+\beta^2_{E_2}(\Phi,\Psi))-\varepsilon,
\end{array}
$$
where $\bar{\rho}=\frac{1}{2}(\rho_1+\rho_2)$ and $\bar{E}=\frac{1}{2}(E_1+E_2)$. This implies the concavity of the function $E\mapsto\beta^2_{E}(\Phi,\Psi)$. The concavity of the function $E\mapsto\beta_{E}(\Phi,\Psi)$ follows from the concavity and monotonicity of the function $\sqrt{x}$.

The last assertion of the proposition follows from (\ref{DB-rel+}) and Proposition 3 in \cite{SCT}, since condition (\ref{H-cond}) guarantees that $H_A$ is an unbounded densely defined operator having  discrete spectrum of finite multiplicity. $\square$

\begin{corollary}\label{ec-b-d-c1} \emph{Let $\Phi$ and $\Psi$ be arbitrary channels from $A$ to $B$ and $R$ be any system. Let $H_A$ be a positive operator on $\mathcal{H}_A$ and $\beta_E$ the corresponding energy-constrained Bures distance defined in (\ref{ec-b-dist}). There exists a common Stinespring representation (\ref{c-S-r})  such that
\begin{equation*}
\hspace{-30pt}\|V_{\Phi}\otimes I_R\, \rho\, V^*_{\Phi}\otimes I_R
-V_{\Psi}\otimes I_R\, \rho\, V^*_{\Psi}\otimes I_R\|_1\displaystyle\leq 2\beta_E(\Phi,\Psi)\leq2\sqrt{\|\Phi-\Psi\|_{\diamond}^{E}}
\end{equation*}
for any state $\rho$ in $\mathfrak{S}(\mathcal{H}_{AR})$ satisfying the inequality $\,\mathrm{Tr} H_A\rho_A\leq E$.}
\end{corollary}\smallskip

\emph{Proof.} Let $\widehat{V}_{\Theta}=V_{\Theta}\otimes I_R$, $\Theta=\Phi,\Psi$. By using Lemma \ref{sl} and the operator Cauchy-Schwarz inequality we obtain
$$
\begin{array}{c}
\|\widehat{V}_{\Phi}\, \rho\, \widehat{V}^*_{\Phi}
-\widehat{V}_{\Psi}\, \rho\, \widehat{V}^*_{\Psi}\|_1\leq 2
\|(\widehat{V}_{\Phi}-\widehat{V}_{\Psi})\rho\|_1=2\mathrm{Tr} W(\widehat{V}_{\Phi}-\widehat{V}_{\Psi})\rho
\\\\\leq 2\sqrt{\mathrm{Tr}(W(\widehat{V}_{\Phi}-\widehat{V}_{\Psi})\rho\shs(\widehat{V}^*_{\Phi}-\widehat{V}^*_{\Psi})W^*)\mathrm{Tr}\rho}=
2\sqrt{\mathrm{Tr}(V_{\Phi}-V_{\Psi})\rho_A(V^*_{\Phi}-V^*_{\Psi})},
\end{array}
$$
where $W$ is the partial isometry from the polar decomposition of $(\widehat{V}_{\Phi}-\widehat{V}_{\Psi})\rho$. So, the assertion of the corollary follows
directly from Proposition \ref{ec-b-d-c1} and  (\ref{DB-rel+}). 

By definition (\ref{c-channel}) the representations (\ref{b-dist}) and  (\ref{ec-b-dist+}) imply the following

\begin{corollary}\label{ec-b-d-c2} \emph{Let $\,\Phi$ and $\,\Psi$ be arbitrary quantum channels from $A$ to $B$.}

A) \emph{There exist\footnote{A complementary channel is defined up to the isometrical equivalence \cite{H-c-ch}.}
 complementary channels $\,\widehat{\Phi}$ and $\,\widehat{\Psi}$ from $A$ to some system $E$ such that $\,\beta(\widehat{\Phi},\widehat{\Psi})\leq\beta(\Phi,\Psi)$.}

B) \emph{Let $\,H_A$ be a positive operator on $\mathcal{H}_A$ and $\beta_E$ the corresponding energy-constrained Bures distance  defined in (\ref{ec-b-dist}). There exist complementary channels $\,\widehat{\Phi}$ and $\,\widehat{\Psi}$ from $A$ to some system $E$ such that $\,\beta_E(\widehat{\Phi},\widehat{\Psi})\leq\beta_E(\Phi,\Psi)$.}
\end{corollary}

\section{Basic lemma}

Our main technical tool is the following lemma proved by modification of the  Alicki-Fannes-Winter method \cite{A&F,W-CB,M&H,AFM}. In this lemma we use the extended QCMI defined in (\ref{cmi-e+}),(\ref{cmi-e++}) and the function $\,g(x)=(1+x)h_2\!\left(\frac{x}{1+x}\right)=(x+1)\log(x+1)-x\log x$.\smallskip

\begin{lemma}\label{S-CMI-CB} \emph{Let $\shs\rho$ and  $\shs\sigma$ be any states in $\,\mathfrak{S}(\mathcal{H}_{ABCDR})$ such that
$\frac{1}{2}\|\shs\rho-\sigma\|_1\leq\varepsilon\leq 1$. Let $\,\mathcal{H}_*$ be a subspace of $\,\mathcal{H}_{AD}$ containing the supports of $\rho_{AD}$ and $\sigma_{AD}$.}

A) \emph{If $\dim\mathcal{H}_*=d<+\infty$ then $I(A\!:\!B|C)_{\rho}$ and $I(A\!:\!B|C)_{\sigma}$  are finite and}
\begin{equation}\label{S-CMI-CB+}
|I(A\!:\!B|C)_{\rho}-I(A\!:\!B|C)_{\sigma}|\leq 2\varepsilon
\log d+2g(\varepsilon).
\end{equation}

\emph{If $\,\rho$ and  $\,\sigma$ are $qc$-states with respect to the decomposition $(AD)(BCR)$ then (\ref{S-CMI-CB+}) holds with the term $\,2\varepsilon
\log d\,$ replaced by $\;\varepsilon
\log d$.}\footnote{$qc$-states are defined in (\ref{chi-rep}); we assume here that the basic $\{|i\rangle\}$ does not depend on a state.}

B) \emph{If $\,\mathrm{Tr} H_*\rho_{AD},\,\mathrm{Tr} H_*\sigma_{AD}\leq E<+\infty$ for some  positive operator $H_*$ on $\mathcal{H}_*$  satisfying condition (\ref{H-cond+}) then $I(A\!:\!B|C)_{\rho}$ and $I(A\!:\!B|C)_{\sigma}$ are finite and
\begin{equation}\label{S-CMI-CB++}
|I(A\!:\!B|C)_{\rho}-I(A\!:\!B|C)_{\sigma}|\leq 2\sqrt{2\varepsilon}
F_{H_{*}}\!\left(E/\varepsilon\right)+2g(\sqrt{2\varepsilon}),
\end{equation}
where $\,F_{H_*}(E)\doteq \sup\{H(\rho)\,|\,\supp \rho\subseteq \mathcal{H}_*, \mathrm{Tr} H_*\rho\leq E\shs\}$.}
\emph{If  $\,\rho\,$ and  $\,\sigma\,$ are pure states then  (\ref{S-CMI-CB++}) holds with $\,\varepsilon\,$  replaced by $\,\varepsilon^2/2$.}

C) \emph{If $\,\rho_{BC}=\sigma_{BC}$ then factor $\,2$ in the last terms in (\ref{S-CMI-CB+}), (\ref{S-CMI-CB++}) and in their specifications for $qc$-states and pure states can be omitted.}
\end{lemma}\smallskip

\emph{Proof.}  A) In the proof of (\ref{S-CMI-CB+}) we may assume that $\,\frac{1}{2}\|\rho-\sigma\|_1=\varepsilon$. Following \cite{W-CB,M&H} introduce the states $\,\gamma_+=\varepsilon^{-1}[\shs\rho-\sigma\shs]_+\,$
and
$\,\gamma_-=\varepsilon^{-1}[\shs\rho-\sigma\shs]_-\,$ in $\mathfrak{S}(\mathcal{H}_{ABCDR})$. Then
\begin{equation*}
\frac{1}{1+\varepsilon}\,\rho+\frac{\varepsilon}{1+\varepsilon}\,\gamma_-=\omega^{*}=
\frac{1}{1+\varepsilon}\,\sigma+\frac{\varepsilon}{1+\varepsilon}\,\gamma_+.
\end{equation*}
By taking partial trace we obtain
\begin{equation*}
\frac{1}{1+\varepsilon}\,\rho_{ABC}+\frac{\varepsilon}{1+\varepsilon}\,[\gamma_-]_{ABC}=\omega^{*}_{ABC}=
\frac{1}{1+\varepsilon}\,\sigma_{ABC}+\frac{\varepsilon}{1+\varepsilon}\,[\gamma_+]_{ABC}.
\end{equation*}
Basic properties of the QCMI and upper bound (\ref{MI-UB}) imply
\begin{equation}\label{s-UB}
I(A\!:\!B|C)_{\omega}\leq I(A\!:\!BC)_{\omega}\leq I(AD\!:\!BCR)_{\omega}\leq 2H(\omega_{AD})
\end{equation}
for any state $\omega$ in $\mathfrak{S}(\mathcal{H}_{ABCDR})$. Since the operators
$\mathrm{Tr}_{BCR}[\rho-\sigma]_{\pm}$ are supported on the subspace $\mathcal{H}_*$, it follows from (\ref{s-UB}) that
\begin{equation}\label{s-UB+}
I(A\!:\!B|C)_{\omega}\leq2\log d<+\infty,\quad \omega=\rho,\sigma,\gamma_+,\gamma_-.
\end{equation}
By applying (\ref{F-c-b}) to the above convex decompositions of $\,\omega_{ABC}^{*}$ we obtain
$$
(1-p)\left[I(A\!:\!B|C)_{\rho}-I(A\!:\!B|C)_{\sigma}\right]\leq p
\left[I(A\!:\!B|C)_{\gamma_+}
-I(A\!:\!B|C)_{\gamma_-}\right]+2\shs
h_2(p),
$$
$$
(1-p)\left[I(A\!:\!B|C)_{\sigma}-I(A\!:\!B|C)_{\rho}\right]\leq p
\left[I(A\!:\!B|C)_{\gamma_-}-
I(A\!:\!B|C)_{\gamma_+}\right]+2\shs h_2(p),
$$
where $p=\frac{\varepsilon}{1+\varepsilon}$. By using  (\ref{s-UB+}) and  nonnegativity of $I(A\!:\!B|C)$ we obtain (\ref{S-CMI-CB+}).

If $\,\rho$ and  $\,\sigma$ are $qc$-states with respect to the decomposition $(AD)(BCR)$ then the
above states $\gamma_+$ and $\gamma_-$ are $qc$-states as well. So, by using (\ref{MI-UB+}) instead of (\ref{MI-UB}) one can remove factor $2$ in (\ref{s-UB}) and (\ref{s-UB+}).

B) We may consider $I(A\!:\!B|C)$ as a function on $\mathfrak{S}(\mathcal{H}_{BCR}\otimes\mathcal{H}_*)$. If $\,\varepsilon<\frac{1}{2}$ then the  assertion can be proved by applying  Proposition 1 in \cite{AFM} to this function by using the nonnegativity of $I(A\!:\!B|C)$, upper bound (\ref{s-UB})
and inequality (\ref{F-c-b}). If $\varepsilon\in[\frac{1}{2},1]$ then the required inequality follows from upper bound (\ref{s-UB}).

C) The possibility to remove factor $2$ from the last term of (\ref{S-CMI-CB++}) in the case $\,\rho_{BC}=\sigma_{BC}$ is shown in \cite[Lemma 2]{UFA}. By inserting the same arguments in the above proof of part A, it is easy to prove the similar assertion for (\ref{S-CMI-CB+}). $\square$

\section{Continuity bounds for the output conditional mutual information}

The quantum mutual information and its conditional version play a basic role in analysis of informational properties of quantum systems and channels (see Sections 6-8).
In this section we  explore continuity properties of the QCMI (defined in (\ref{cmi-e+}),(\ref{cmi-e++})) at the output of a channel acting on one subsystem of a tripartite system, i.e. the quantity $I(B\!:\!D|C)_{\Phi\otimes\id_{CD}(\rho)}$, where $\Phi:A\rightarrow B$ is an arbitrary channel, $C,D$ are any systems and $\rho$ is a state in $\mathfrak{S}(\mathcal{H}_{ADC})$.
We will obtain continuity bounds for the function $$(\Phi, \rho)\mapsto I(B\!:\!D|C)_{\Phi\otimes\id_{CD}(\rho)}$$
assuming that the set of all channels from $A$ to $B$ is equipped with the Bures distance (\ref{b-dist}) in the case $\dim\mathcal{H}_A<+\infty$  and with the energy-constrained Bures distance (\ref{ec-b-dist}) in the case $\dim\mathcal{H}_A=+\infty$. We will also obtain continuity bounds for the function
$$
\Phi\mapsto I(B^n\!:\!D|C)_{\Phi^{\otimes n}\otimes\id_{CD}(\rho)}
$$
for arbitrary $\rho\in\mathfrak{S}(\mathcal{H}_{A^nCD})$ and any natural $\shs n$.

\subsection{QCMI at the output of a local channel}

\subsubsection{Finite input dimension} The following proposition contains continuity bounds for the function $(\Phi, \rho)\mapsto I(B\!:\!D|C)_{\Phi\otimes\id_{CD}(\rho)}$, where $\Phi$ runs over all channels with a finite-dimensional input system $A$ and $\rho$ runs over all states of a tripartite system $ACD$.

\begin{property}\label{MI-CB} \emph{Let $\,\Phi$ and $\,\Psi$ be quantum channels from a finite-dimensional system $A$ to arbitrary system $B$, $C$  and $D$ be any systems, $\,d_A\doteq\dim\mathcal{H}_A$. Then
\begin{equation}\label{MI-CB+}
\hspace{-30pt}|I(B\!:\!D|C)_{\Phi\otimes\id_{CD}(\rho)}-I(B\!:\!D|C)_{\Psi\otimes\id_{CD}(\sigma)}|\leq 2\varepsilon\log d_A+2\varepsilon
\log 2+2g(\varepsilon),\!
\end{equation}
for any states $\rho$ and $\sigma$ in  $\,\mathfrak{S}(\mathcal{H}_{ACD})$, where $\,\varepsilon=\frac{1}{2}\|\shs\rho-\sigma\|_1+\beta(\Phi,\Psi)$.}

\emph{If $\,\Phi=\Psi$ then the term $\shs2\varepsilon
\log 2$ in (\ref{MI-CB+}) can be omitted. If $\,\rho=\sigma$ then (\ref{MI-CB+}) holds without factor  $2$ in the last term.}

\emph{Continuity bound (\ref{MI-CB+}) is tight in  both cases $\,\Phi=\Psi$ and $\,\rho=\sigma$ for any system $C$. The Bures distance $\beta(\Phi,\Psi)$ in (\ref{MI-CB+}) can be replaced by $\sqrt{\|\Phi-\Psi\|_{\diamond}}$.}
\end{property}\smallskip

\emph{Proof.} By Theorem 1 in \cite{Kr&W} there is a common Stinespring representation (\ref{c-S-r}) such that $\|V_{\Phi}-V_{\Psi}\|=\beta(\Phi,\Psi)$.
Then $\,\hat{\rho}=V_{\Phi}\otimes I_{CD}\shs\rho\shs V_{\Phi}^*\otimes I_{CD}\,$
and $\,\hat{\sigma}=V_{\Psi}\otimes I_{CD}\shs\sigma\shs V_{\Psi}^*\otimes I_{CD}\,$
are extensions of the states $\Phi\otimes\id_{CD}(\rho)$
and $\Psi\otimes\id_{CD}(\sigma)$. Lemma \ref{sl} implies
\begin{equation}\label{norm}
 \|\shs\hat{\rho}-\hat{\sigma}\|_1\leq \|\shs\rho-\sigma\|_1+2\|V_{\Phi}-V_{\Psi}\|.
\end{equation}
Since the states  $\hat{\rho}_{BE}=V_{\Phi}\rho_AV_{\Phi}^*$ and $\hat{\sigma}_{BE}=V_{\Psi}\sigma_AV_{\Psi}^*$
are supported on the subspace $V_{\Phi}\mathcal{H}_A\vee V_{\Psi}\mathcal{H}_A$ of $\mathcal{H}_{BE}$ whose dimension does not exceed $2d_A$,
Lemma \ref{S-CMI-CB}A and inequality (\ref{norm}) imply (\ref{MI-CB+}). If $\Phi=\Psi$ then the above states
$\hat{\rho}_{BE}$ and $\hat{\sigma}_{BE}$ are supported on the $d_A$-dimensional subspace  $V_{\Phi}\mathcal{H}_A=V_{\Psi}\mathcal{H}_A$.
If $\,\rho=\sigma$ then $\hat{\rho}_{CD}=\hat{\sigma}_{CD}$ and (\ref{MI-CB+}) holds with $2g(\varepsilon)$ replaced by $g(\varepsilon)$ by Lemma \ref{S-CMI-CB}C.

The tightness of continuity bound (\ref{MI-CB+}) in the case $\Phi=\Psi$ follows from Corollary 2 in \cite{CHI}.

The tightness of continuity bound (\ref{MI-CB+}) in the case $\rho=\sigma$ can be derived from the tightness of continuity bound (\ref{QC-CB}) for the quantum capacity. It can be also shown directly by using the erasure channels $\Phi_{1/2}$ and $\Phi_{1/2-x}$ (see the proof of Theorem \ref{cap-tcb} in Section 7).

The last assertion of the proposition follows from the right inequality in (\ref{DB-rel}) and monotonicity of the function $g(x)$.$\square$

\subsubsection{Finite input energy}

Note first that the (uniform) continuity bound for the function $(\Phi, \rho)\mapsto I(B\!:\!D|C)_{\Phi\otimes\id_{CD}(\rho)}$
under the energy constraint on $\rho_A$ can be obtained by combining
(uniform) continuity bounds for the functions $\,\rho\mapsto I(B\!:\!D|C)_{\Phi\otimes\id_{CD}(\rho)}\,$ and
 $\,\Phi\mapsto I(B\!:\!D|C)_{\Phi\otimes\id_{CD}(\rho)}\,$ under this constraint \emph{not depending} on $\Phi$ and $\rho$.

\begin{property}\label{CMI-fe} \emph{Let $\,\Phi:A\rightarrow B$ be an arbitrary quantum channel, $C$  and $D$  any systems.
If the Hamiltonian $H_{A}$ of the input system $A$ satisfies condition (\ref{H-cond+}) then the function
$\,\rho_{ACD}\mapsto I(B\!:\!D|C)_{\Phi\otimes\id_{CD}(\rho)}$ is uniformly continuous on the set of states with bounded energy of $\,\rho_{A}$. Quantitatively,
\begin{equation}\label{CMI-CB-fe}
\hspace{-30pt}|I(B\!:\!D|C)_{\Phi\otimes\id_{C}(\rho)}-I(B\!:\!D|C)_{\Phi\otimes\id_{C}(\sigma)}|\leq 2\sqrt{2\varepsilon} \bar{F}_{H_{A}}\!\left(\bar{E}/\varepsilon\right)+2g(\sqrt{2\varepsilon})
\end{equation}
for any states $\rho$ and $\sigma$ in $\mathfrak{S}(\mathcal{H}_{ACD})$ such that $\mathrm{Tr} H_{A}\rho_{A},\mathrm{Tr} H_{A}\sigma_{A}\leq E$ and   $\;\frac{1}{2}\|\shs\rho-\sigma\|_1\leq\varepsilon$, where $\,\bar{F}_{H_{A}}\,$ is the function defined in (\ref{F-bar}) and $\,\bar{E}=E-E_0$.}

\emph{If $\shs\rho\shs$ and  $\shs\sigma\shs$ are  pure states then  (\ref{CMI-CB-fe}) holds with $\,\varepsilon$ replaced by $\,\varepsilon^2/2$.}

\emph{If $\,A$ is the $\ell$-mode quantum oscillator then the function $\bar{F}_{H_A}$ in (\ref{CMI-CB-fe}) can be replaced by its upper bound $\bar{F}_{\ell,\omega}$ defined in (\ref{F-ub+})}.
\end{property}\smallskip

The main term in (\ref{CMI-CB-fe}) tends to zero as $\,\varepsilon\!\rightarrow\!0^+$, since condition (\ref{H-cond+}) implies that $\,\bar{F}_{H_{A}}(E)=o\shs(\sqrt{E})\,$ as $\,E\rightarrow\!+\infty$ (see Section 2.2.1).\smallskip

\emph{Proof.} Assume that the channel $\Phi$ has the Stinespring representation (\ref{St-rep}). Continuity bound (\ref{CMI-CB-fe}) follows from Lemma \ref{S-CMI-CB}B with $\mathcal{H}_*=V_{\Phi}\mathcal{H}_{A}\subseteq\mathcal{H}_{BE}$ and  $H_*=V_{\Phi}H_{A}V_{\Phi}^*-E_0I_{\mathcal{H}_*}$. $\square$

The continuity bound for the function $\,\Phi\mapsto I(B\!:\!D|C)_{\Phi\otimes\id_{CD}(\rho)}\,$
under the energy constraint on $\rho_A$ not depending  on $\Phi$ and $\rho$ is presented in Proposition \ref{omi-fe} below (the case $n=1$).

\subsection{QCMI at the output of n copies of a local channel}

\subsubsection{Finite input dimension}

The following proposition contains continuity bound for the function $\Phi\mapsto I(B^n\!:\!D|C)_{\Phi^{\otimes n}\otimes\id_{CD}(\rho)}$,
where $\Phi$ runs over all channels with a finite-dimensional input system $A$ and $\rho$ is a fixed state of a tripartite system $ACD$.
It is proved by using the Leung-Smith telescopic method from \cite{L&S}.

\begin{property}\label{omi} \emph{Let $\,\Phi$ and $\,\Psi$ be quantum channels from a finite-dimensional system $A$ to arbitrary system $B$, $C$ and $D$ be any systems and $\,n\in\mathbb{N}$. Then
\begin{equation}\label{omi+}
\hspace{-45pt}\left|I(B^n\!:\!D|C)_{\Phi^{\otimes n}\otimes\id_{CD}(\rho)}-I(B^n\!:\!D|C)_{\Psi^{\otimes n}\otimes\id_{CD}(\rho)}\right|\leq n(2\varepsilon
\log (2d_A) + g(\varepsilon))
\end{equation}
for any state $\,\rho\shs$ in $\,\mathfrak{S}(\mathcal{H}^{\otimes n}_{A}\otimes\mathcal{H}_{CD})$, where $\,\varepsilon=\beta(\Phi,\Psi)$ and $\,d_A=\dim\mathcal{H}_A$.}

\emph{Continuity bound (\ref{omi+}) is tight for any system $C$ (for each given $n$ and large $d_A$). The Bures distance $\beta(\Phi,\Psi)$ in (\ref{omi+}) can be replaced by $\sqrt{\|\Phi-\Psi\|_{\diamond}}$.}
\end{property}\smallskip

\emph{Proof.} Consider the states $\sigma_k=\Phi^{\otimes k}\otimes\Psi^{\otimes (n-k)}\otimes\id_{CD}(\rho)$, $k=0,1,...,n$. Then
\begin{equation}\label{tel}
\hspace{-45pt}\begin{array}{c}
\displaystyle \left|I(B^n\!:\!D|C)_{\sigma_n}\!-I(B^n\!:\!D|C)_{\sigma_0}\right|\displaystyle=
\left|\sum_{k=1}^n I(B^n\!:\!D|C)_{\sigma_k}\!-I(B^n\!:\!D|C)_{\sigma_{k-1}}\right|\\ \leq \displaystyle \sum_{k=1}^n \left|I(B^n\!:\!D|C)_{\sigma_k}\!-I(B^n\!:\!D|C)_{\sigma_{k-1}}\right|.
\end{array}\!\!\!
\end{equation}
For each $k$ the chain rule (\ref{chain}) implies that
\begin{equation}\label{tel+}
\hspace{-50pt}\begin{array}{l}
I(B^n\!:\!D|C)_{\sigma_k}\!-I(B^n\!:\!D|C)_{\sigma_{k-1}} = I(Y_k \!:\!D|C)_{\sigma_k} +\,I(B_k\!:\!D|Y_kC)_{\sigma_k}\\\\-\;
I(Y_k \!:\!D|C)_{\sigma_{k-1}}-\,I(B_k\!:\!D|Y_kC)_{\sigma_{k-1}}=
I(B_k\!:\!D|Y_kC)_{\sigma_k}-
I(B_k\!:\!D|Y_kC)_{\sigma_{k-1}},
\end{array}\!\!
\end{equation}
where $Y_k=B_1..B_{k-1}B_{k+1}..B_n$. The second equality here follows from the equality $\mathrm{Tr}_{B_k}\sigma_k=\mathrm{Tr}_{B_k}\sigma_{k-1}$. Note that the finite entropy of the states $\,\rho_{A_1},...,\rho_{A_n}$, upper bound (\ref{CMI-UB}) and monotonicity of the QCMI under local channels guarantee finiteness of all the terms in (\ref{tel}) and (\ref{tel+}).

By Theorem 1 in \cite{Kr&W} there is a common Stinespring representation (\ref{c-S-r}) such that $\|V_{\Phi}-V_{\Psi}\|=\beta(\Phi,\Psi)$.
To estimate the last difference in (\ref{tel+}) consider  the states
\begin{equation}\label{s-one}
\hat{\sigma}_k= W_k\otimes V^k_{\Phi}\otimes I_{CD} \cdot\rho\cdot W^*_k\otimes [V^k_{\Phi}]^*\otimes I_{CD},
\end{equation}
\begin{equation}\label{s-two}
\hat{\sigma}_{k-1}= W_k\otimes V^k_{\Psi}\otimes I_{CD} \cdot\rho\cdot W^*_k\otimes [V^k_{\Psi}]^*\otimes I_{CD}
\end{equation}
in $\mathfrak{S}(\mathcal{H}_{B^nE^nCD})$, where $\,W_k=V^1_{\Phi}\otimes\ldots \otimes V^{k-1}_{\Phi}\otimes V^{k+1}_{\Psi}\otimes\ldots \otimes V^{n}_{\Psi}\,$ is an isometry from $\mathcal{H}_{A^n\setminus A_k}$ into $\mathcal{H}_{[BE]^n\setminus [BE]_k}$, $V^k_{\Phi}\cong V_{\Phi}$ and $V^k_{\Psi}\cong V_{\Psi}$ are isometries from $\mathcal{H}_{A_k}$ into $\mathcal{H}_{B_kE_k}$. It follows from (\ref{c-S-r}) that these states
are extensions of the states $\sigma_k$ and $\sigma_{k-1}$, i.e. $\mathrm{Tr}_{E^n}\hat{\sigma}_k=\sigma_k$ and $\mathrm{Tr}_{E^n}\hat{\sigma}_{k-1}=\sigma_{k-1}$. Since
$[\hat{\sigma}_k]_{B_kE_k}=V^k_{\Phi}\rho_{A_k}[V^k_{\Phi}]^*$, $[\hat{\sigma}_{k-1}]_{B_kE_k}=V^k_{\Psi}\rho_{A_k} [V^k_{\Psi}]^*$ and $\,\mathrm{Tr}_{B_kE_k}\hat{\sigma}_{k}=\mathrm{Tr}_{B_kE_k}\hat{\sigma}_{k-1}$, Lemma \ref{S-CMI-CB}C implies that
\begin{equation}\label{tel++}
|I(B_k\!:\!D|Y_kC)_{\sigma_k}-
I(B_k\!:\!D|Y_kC)_{\sigma_{k-1}}|\leq 2\varepsilon' \log (2\shs d_A)+g(\varepsilon'),
\end{equation}
where $\,\varepsilon'=\frac{1}{2}\|\hat{\sigma}_{k}-\hat{\sigma}_{k-1}\|_{1}$. By using Lemma \ref{sl} we obtain
$$
\varepsilon' \leq\|W_k\otimes V^k_{\Phi}\otimes I_{CD}-W_k\otimes V^k_{\Psi}\otimes I_{CD}\|=\|V_{\Phi}-V_{\Psi}\|=\beta(\Phi,\Psi)=\varepsilon.
$$
Hence, it follows from  (\ref{tel+}) and (\ref{tel++}) that $|I(B^n\!:\!D|C)_{\sigma_k}-I(B^n\!:\!D|C)_{\sigma_{k-1}}|$ does not exceed   $2\varepsilon \log (2\shs d_A)+g(\varepsilon)$. So, (\ref{tel}) implies the required inequality.

The tightness of the continuity bound in Proposition \ref{omi} follows from the tightness of continuity bound (\ref{MI-CB+}) in the case $\,\rho=\sigma$, since
for arbitrary channel $\Phi:A\rightarrow B$, any system $D$ and a state $\rho\in\mathfrak{S}(\mathcal{H}_{AD})$ we have
$I(B^n\!:\!D^n)_{\Phi^{\otimes n}\otimes\id_{D^n}(\rho^{\otimes n})}=nI(B\!:\!D)_{\Phi\otimes\id_{D}(\rho)}$.

The last assertion of the proposition follows from the right inequality in (\ref{DB-rel}) and monotonicity of the function $g(x)$.
$\square$

\subsubsection{Finite input energy}

In this subsection we assume that $H_A$ is  the Hamiltonian of a system $A$ satisfying condition (\ref{H-cond+}).
We will use the function $\,\bar{F}_{H_A}$ defined in (\ref{F-bar}) and
the increasing sequence $\,\{\gamma(d)=\bar{F}^{-1}_{H_A}(\log d)\}_{d\geq d_0}$ introduced in Lemma \ref{cl} tending to $+\infty$ as $d\rightarrow+\infty$, where $d_0$ is the multiplicity of the minimal eigenvalue $E_0$ of $H_A$.
We will also use the energy-constrained Bures distance $\beta_E$ between quantum channels and the energy-constrained diamond norm $\,\|\cdot\|_{\diamond}^E$ defined, respectively, in (\ref{ec-b-dist}) and (\ref{ecd}).

\begin{property}\label{omi-fe} \emph{Let $\,\Phi$ and $\,\Psi$ be arbitrary quantum channels from $A$ to $B$, $C$ and $D$  any systems, $\,n\in\mathbb{N}$ and $\,\rho\shs$  a state in $\,\mathfrak{S}(\mathcal{H}^{\otimes n}_{A}\otimes\mathcal{H}_{CD})$ such that $\,\sum_{k=1}^n\mathrm{Tr} H_A\rho_{A_k}\leq nE $. Let $s=0$ if the function $\,E\mapsto \bar{F}_{H_A}(E)/\sqrt{E}\,$ is non-increasing on $\,\mathbb{R}_+$ and $\,s=1$ otherwise. Let $\,t=0\,$ if  $\,\mathrm{Tr} H_A\rho_{A_k}\leq E$ for all $\,k$ and  $\,t=1$ otherwise.  Then
\begin{equation}\label{omi-fe-cb}
\hspace{-75pt}\begin{array}{c}
\left|I(B^n\!\!:\!D|C)_{\Phi^{\otimes n}\otimes\id_{CD}(\rho)}-I(B^n\!\!:\!D|C)_{\Psi^{\otimes n}\otimes\id_{CD}(\rho)}\right|
\!\leq n(T_{s,t}(E,\varepsilon)+g(\varepsilon)+2\varepsilon\log 2),
\end{array}
\end{equation}
where $\,\varepsilon=\beta_E(\Phi,\Psi)\,$ and 
\begin{equation}\label{omi-c}
\hspace{-30pt}T_{s,t}(E,\varepsilon)\doteq\!\min_{\gamma(d)\geq2\bar{E}} \left[\left(\!4\sqrt{\frac{2^{s}\bar{E}}{\gamma(d)}}+\frac{4st\bar{E}}{\gamma(d)}+2\varepsilon\!\right)\log d+4g\!\left(\!\sqrt{\frac{2^s\bar{E}}{\gamma(d)}}\right)\right].
\end{equation}
(the minimum here is over natural numbers $d$ such that $\,\frac{1}{2}\gamma(d)\geq\bar{E}\doteq E-E_0$).}

\emph{Inequality (\ref{omi-fe-cb}) also holds with $\,\varepsilon=\sqrt{\|\Phi-\Psi\|^{E}_{\diamond}}$.}\smallskip

\emph{For given $\,E\geq E_0$, $\,s$ and $\,t$ the quantity  $T_{s,t}(E,\varepsilon)$ tends to zero as $\,\varepsilon\rightarrow0^+$.}\smallskip
 \end{property}

\begin{remark}\label{omi-fe-r}
The quantity $T_{s,t}(E,\varepsilon)$ is determined by the function
$\bar{F}_{H_A}$ (since $\gamma(d)$ is determined by $\bar{F}_{H_A}$), i.e by the Hamiltonian $H_A$ of the system $A$. It can be calculated for any $\varepsilon>0\shs$ and $E>E_0$ by using explicit expression for $\bar{F}_{H_A}$. If this expression is not known (or too complicated) we can use any upper bound $\widehat{F}_{H_A}$ for $\bar{F}_{H_A}$ provided that
$\widehat{F}_{H_A}$ is a concave nonnegative function on $\,[0,+\infty)\,$ such that $\widehat{F}_{H_A}(E)=o\shs(\sqrt{E})$ as $E\rightarrow+\infty$. It  follows from the proofs of Proposition
\ref{omi-fe} and Lemma \ref{cl} that \emph{the quantity $T_{s,t}(E,\varepsilon)$ in Proposition
\ref{omi-fe} can be determined by formula (\ref{omi-c}) with $\gamma(d)$ and $s$  replaced, respectively, by  $\hat{\gamma}(d)\doteq\widehat{F}_{H_A}^{-1}(\log d)$ and the corresponding $\,\hat{s}$.}
This approach will be used to specify  Proposition
\ref{omi-fe} for the case when $A$ is a multi-mode quantum oscillator (see Corollary \ref{omi-fe-c} below).
\end{remark}

\emph{Proof of Proposition \ref{omi-fe}.} We can repeat the arguments from the proof of Proposition \ref{omi} up to  the estimation of the last difference in (\ref{tel+}). In the case $st=1$ we have to estimate this difference by two different ways depending on the value of $\,x_k\doteq\mathrm{Tr} H_A\rho_{A_k}$.

Let $d$ be a natural number such that $\gamma(d)\geq 2\bar{E}$,  $N_1$ the set of all indexes $k$ such that  $\,\bar{x}_k\doteq x_k-E_0\leq\gamma(d)/2\,$ and $N_2=\{1,...,n\}\setminus N_1$. Let $n_i=\sharp (N_i)$,  $X_i=n_i^{-1}\sum_{k\in N_i}x_k$ and $\bar{X}_i=X_i-E_0$, $i=1,2$. Let $\hat{\rho}$ be a pure state in $\mathfrak{S}(H_{A^nCDR})$ such that $\mathrm{Tr}_R\hat{\rho}=\rho$.

Assume first that $k\in N_1$. Then $\Delta^k_d\doteq\sqrt{2\bar{x}_k/\gamma(d)}\leq 1$. Lemma \ref{cl} implies existence of a pure state $\varrho^k$ in $\,\mathfrak{S}(\mathcal{H}_{A^nCDR})$ such that
$\rank \varrho^k_{A_k}\leq d$, $\,\mathrm{Tr} H_A\varrho^k_{A_k}\leq x_k$, $\,\textstyle\frac{1}{2}\|\hat{\rho}-\varrho^k\|_1\leq \sqrt{\bar{x}_k/\gamma(d)}\,$ and
\begin{equation}\label{vr}
\|\hat{\rho}-\varrho^k\|_1 \mathrm{Tr} \bar{H}_A[[\hat{\rho}-\varrho^k]_{\pm}]_{A_k}\leq 2\bar{x}_k, \quad \bar{H}_A=H_A-E_0I_A.
\end{equation}

By Corollary \ref{ec-b-d-c1} for each $k$ there exists a common Stinespring representation (\ref{c-S-r}) with the isometries $V^k_{\Phi}$ and $V^k_{\Psi}$ from $\mathcal{H}_A$ into $\mathcal{H}_{BE_k}$ such that
$$
\textstyle\frac{1}{2}\|V^k_{\Phi}\otimes I_{T_k}\, \varrho^k\, [V^k_{\Phi}]^*\otimes I_{T_k}
-V^k_{\Psi}\otimes I_{T_k}\, \varrho^k\, [V^k_{\Psi}]^*\otimes I_{T_k}\|_1\leq \varepsilon_k\doteq\beta_{x_k}(\Phi,\Psi),
$$
where $T_k=A^nCDR\setminus A_k$. Consider the pure states
\begin{equation}\label{s-one+}
\hat{\varsigma}^k_k= W_k\otimes V^k_{\Phi}\otimes I_{CDR} \cdot\varrho^k\cdot W^*_k\otimes [V^k_{\Phi}]^*\otimes I_{CDR},
\end{equation}
\begin{equation}\label{s-two+}
\hat{\varsigma}^k_{k-1}= W_k\otimes V^k_{\Psi}\otimes I_{CDR} \cdot\varrho^k\cdot W^*_k\otimes [V^k_{\Psi}]^*\otimes I_{CDR}
\end{equation}
in $\mathfrak{S}(\mathcal{H}_{B^nE_1...E_nCDR})$, where $\,W_k$ is the same isometry as in (\ref{s-one}),(\ref{s-two}), while  $V^k_{\Phi}$, $V^k_{\Psi}$ are the isometries chosen before. Then
$\|\hat{\varsigma}^k_k-\hat{\varsigma}^k_{k-1}\|_1\leq \varepsilon_k$. Since $\rank \varrho^k_{A_k}\leq d$, Lemma \ref{S-CMI-CB}C implies that
\begin{equation*}
\hspace{-30pt}|I(B_k\!:\!D|Y_kC)_{\hat{\varsigma}^k_k}-
I(B_k\!:\!D|Y_kC)_{\hat{\varsigma}^k_{k-1}}|\leq  2\varepsilon_k \log (2d)+g(\varepsilon_k)\qquad   (Y_k=B^n\setminus B_{k}).
\end{equation*}

Let $\hat{\sigma}_k$ and $\hat{\sigma}_{k-1}$ be the pure states in $\,\mathfrak{S}(\mathcal{H}_{B^nE_1...E_nCDR})$ defined, respectively, by formulas (\ref{s-one+}) and (\ref{s-two+}) with $\varrho^k$ replaced by $\hat{\rho}$. Since   $\,\mathrm{Tr} H_A\varrho^k_{A_k},\mathrm{Tr} H_A\rho_{A_k}\leq x_k$ and $\,\|\hat{\sigma}_{k-1}-\hat{\varsigma}^k_{k-1}\|_1=\|\hat{\sigma}_k-\hat{\varsigma}^k_k\|_1=\|\hat{\rho}-\varrho^k\|_1\leq2\Delta^k_d\leq 2\,$,  Lemma \ref{S-CMI-CB}B with $\,\mathcal{H}_*=V^k_{\Theta}\mathcal{H}_{A}\subseteq\mathcal{H}_{B_kE_k}\,$ and  $\,H_*=V^k_{\Theta}H_{A}[V^k_{\Theta}]^*-E_0I_{\mathcal{H}_*}$, $\,\Theta=\Phi,\Psi$,  implies that
\begin{equation}\label{quant}
\hspace{-70pt}|I(B_k\!:\!D|Y_kC)_{\sigma_{k-1}}-
I(B_k\!:\!D|Y_kC)_{\hat{\varsigma}^k_{k-1}}|\;\;\textrm{and}\;\;|I(B_k\!:\!D|Y_kC)_{\sigma_k}-
I(B_k\!:\!D|Y_kC)_{\hat{\varsigma}^k_{k}}|
\end{equation}
are bounded above by
\begin{equation}\label{tmp-ub+}
2\Delta^k_d\bar{F}_{H_{A}}\!\left(\frac{2\bar{x}_k}{(\Delta^k_d)^2}\right)+2g(\Delta^k_d)
= 2\sqrt{\frac{2\bar{x}_k}{\gamma(d)}}\log d+2g\!\left(\!\sqrt{\frac{2\bar{x}_k}{\gamma(d)}}\right),
\end{equation}
where the equality $\,\bar{F}_{H_{A}}(\gamma(d))=\log d\,$ was used.

Thus, the last difference in (\ref{tel+}) is bounded above by
$$
B^1_k(d)\doteq4\sqrt{\frac{2\bar{x}_k}{\gamma(d)}}\log d+4g\!\left(\!\sqrt{\frac{2\bar{x}_k}{\gamma(d)}}\right)+2\varepsilon_k \log (2d)+g(\varepsilon_k).
$$
Let $\bar{\varepsilon}=n_1^{-1}\sum_{k\in N_1}\varepsilon_k$. By using the concavity of the functions $\sqrt{x}$ and $g(x)$ along with the monotonicity of $g(x)$, we obtain
\begin{equation}\label{s-1}
\hspace{-30pt}\sum_{k\in N_1}B^1_k(d)\leq
4n_1\sqrt{\frac{2\bar{X}_1}{\gamma(d)}}\log d+4g\!\left(\!\sqrt{\frac{2\bar{X}_1}{\gamma(d)}}\right)
 +2\bar{\varepsilon}n_1\log (2d)+n_1g(\bar{\varepsilon}),
\end{equation}

For each $k\in N_2$ let $\hat{\sigma}_k$ and $\hat{\sigma}_{k-1}$ be the states in $\,\mathfrak{S}(\mathcal{H}_{B^nE_1...E_nCD})$ defined, respectively, by formulas (\ref{s-one}) and (\ref{s-two}) with arbitrary sets $\{V^j_{\Phi}\}$ and $\{V^j_{\Psi}\}$ of  isometries from common Stinespring representation (\ref{c-S-r}). Since $I(B_k\!:\!D|Y_kC)\leq I(B_kE_k\!:\!D|Y_kC)$, upper bound (\ref{CMI-UB}) implies
that in this case the last difference in (\ref{tel+}) is bounded above by
$$
B^2_k\doteq \max\shs\{I(B_kE_k\!:\!D|Y_kC)_{\hat{\sigma}_k}, I(B_kE_k\!:\!D|Y_kC)_{\hat{\sigma}_{k-1}}\}\leq 2 H(\rho_{A_k})\leq 2F_{H_{A}}(x_k).
$$
Since $(n-n_2)X_1+n_2X_2\leq nE$ and $X_1\geq E_0$, we have $X_2\leq n\bar{E}/n_2+E_0$. So, by using concavity and monotonicity of the function $\,F_{H_A}$ we obtain
\begin{equation}\label{s-2}
\sum_{k\in N_2}B^2_k\leq 2\sum_{k\in N_2} F_{H_{A}}(x_k)\leq 2n_2F_{H_{A}}(X_2)\leq 2n_2\bar{F}_{H_{A}}(n\bar{E}/n_2).
\end{equation}
It is easy to see that $X_1\leq E$. So, since the function $E\mapsto \beta_{E}(\Phi,\Psi)$ is concave and nondecreasing by Proposition \ref{ec-b-d-c1}, we have
\begin{equation}\label{eps-ub}
  \bar{\varepsilon}=\frac{1}{n_1}\sum_{k\in N_1}\beta_{x_k}(\Phi,\Psi)\leq\beta_{X_1}(\Phi,\Psi)\leq\beta_{E}(\Phi,\Psi)=\varepsilon.
\end{equation}
Since $\bar{x}_k>\gamma(d)/2$ for all $k\in N_2$ and $(n-n_2)E_0+\sum_{k\in N_2}\bar{x}_k+n_2E_0\leq\sum_{k\in N_1}x_k+\sum_{k\in N_2}x_k\leq nE$, we have $n_2/n\leq 2\bar{E}/\gamma(d)$. So, it follows from (\ref{tel}),(\ref{tel+}),(\ref{s-1})-(\ref{eps-ub}) and Lemma \ref{GWL} that the l.h.s. of (\ref{omi-fe-cb}) is bounded above by
\begin{equation}\label{t-ub-1}
\hspace{-30pt}\begin{array}{rl}
\displaystyle\sum_{k\in N_1}B^1_k(d)+\sum_{k\in N_2}B^2_k\!&\leq \displaystyle\; n\left(4\sqrt{\frac{2\bar{E}}{\gamma(d)}}+2\varepsilon\right)\log d+4ng\!\left(\!\sqrt{\frac{2\bar{E}}{\gamma(d)}}\right)\\\\ &\displaystyle\;+\;2n\varepsilon\log 2+ng(\varepsilon)+\frac{4n\bar{E}}{\gamma(d)}\bar{F}_{H_{A}}\!\!\left(\frac{\gamma(d)}{2}\right).
\end{array}
\end{equation}
Since $\bar{F}_{H_{A}}(\gamma(d)/2)\leq\bar{F}_{H_{A}}(\gamma(d))=\log d$, it implies (\ref{omi-fe-cb}).

If $\,t=0$, i.e. $\mathrm{Tr} H_A\rho_{A_k}\leq E$ for all $k$
then the above-defined set $N_2$ is empty for all $\,d\,$ such that $\,\gamma(d)\geq2\bar{E}$. So, the last term in (\ref{t-ub-1}) disappears
in this case.

If $\,s=0$, i.e. the function $\,E\mapsto \bar{F}_{H_A}(E)/\sqrt{E}\,$ is non-increasing then we can
make more subtle estimate of the quantities in (\ref{quant}) by using property (\ref{vr}).

Let $\alpha_k=\frac{1}{2}\|\hat{\rho}-\varrho^k\|_1$. Consider the states $\gamma^k_{-}=\alpha^{-1}_k[\hat{\sigma}_k-\hat{\varsigma}^k_k]_{-}$ and $\gamma^k_{+}=\alpha^{-1}_k[\hat{\sigma}_k-\hat{\varsigma}^k_k]_{+}$ which are determined by formula (\ref{s-one+}) with $\varrho^k$ replaced, respectively,
by $\alpha^{-1}_k[\hat{\rho}-\varrho^k]_{-}$ and $\alpha^{-1}_k[\hat{\rho}-\varrho^k]_{+}$. We have
\begin{equation*}
\frac{1}{1+\alpha_k}\,\hat{\sigma}_k+\frac{\alpha_k}{1+\alpha_k}\,\gamma^k_-=\omega^{*}=
\frac{1}{1+\alpha_k}\,\hat{\varsigma}^k_k+\frac{\alpha_k}{1+\alpha_k}\,\gamma^k_+.
\end{equation*}
By using the arguments from the proof of Lemma \ref{S-CMI-CB}A (based on relation (\ref{F-c-b})) we obtain that
the quantity $\left|I(B_k\!:\!D|Y_kC)_{\hat{\sigma}_k}-I(B_k\!:\!D|Y_kC)_{\hat{\varsigma}^k_k}\right|$ is bounded above by
\begin{equation}\label{s-ineq}
\alpha_k\max\!\left\{I(B_k\!:\!D|Y_kC)_{\gamma^k_-},I(B_k\!:\!D|Y_kC)_{\gamma^k_+}\right\}+2g(\alpha_k)
\end{equation}
Basic properties of the QCMI and upper bound (\ref{CMI-UB}) imply
\begin{equation*}
\hspace{-30pt}I(B_k\!:\!D|Y_kC)_{\gamma^k_\pm}\leq I(B_kE_k\!:\!D|Y_kC)_{\gamma^k_\pm}\leq  2H([\gamma^k_\pm]_{B_kE_k})
\leq 2H\!\left(\alpha^{-1}_k[[\hat{\rho}-\varrho^k]_{\pm}]_{A_k}\right).
\end{equation*}
It follows from (\ref{vr}) that $H(\alpha^{-1}_k[[\hat{\rho}-\varrho^k]_{\pm}]_{A_k})\leq F_{\bar{H}_A}(\bar{x}_k/\alpha^2_k)=\bar{F}_{H_A}(\bar{x}_k/\alpha^2_k)$.
So, (\ref{s-ineq}) is bounded above by
\begin{equation}\label{new-est}
$$
\begin{array}{c}
\displaystyle 2\alpha_k\bar{F}_{H_A}\left(\frac{\bar{x}_k}{\alpha^2_k}\right)+2g(\alpha_k)\leq 2\sqrt{\frac{\bar{x}_k}{\gamma(d)}}\bar{F}_{H_A}(\gamma(d))+2g\!\left(\!\sqrt{\frac{\bar{x}_k}{\gamma(d)}}\right)\\\\
=\displaystyle  2\sqrt{\frac{\bar{x}_k}{\gamma(d)}}\log d+2g\!\left(\!\sqrt{\frac{\bar{x}_k}{\gamma(d)}}\right).
\end{array}
$$
\end{equation}
Since $\alpha_k\leq\sqrt{\bar{x}_k/\gamma(d)}$, the inequality in (\ref{new-est}) follows from the assumed non-increasing property of the function $\,E\mapsto \bar{F}_{H_A}(E)/\sqrt{E}\,$
and the non-decreasing property of $g$. The equality follows from the definition of $\gamma(d)$.

Similar arguments show that the quantity $\left|I(B_k\!:\!D|Y_kC)_{\hat{\sigma}_{k-1}}-I(B_k\!:\!D|Y_kC)_{\hat{\varsigma}^k_{k-1}}\right|$
is also bounded above by (\ref{new-est}).

If $\,\bar{x}_k>\gamma(d)$ then $\sqrt{\bar{x}_k/\gamma(d)}\bar{F}_{H_A}(\gamma(d))\geq \bar{F}_{H_A}(\bar{x}_k)=F_{H_A}(x_k)$. So, (\ref{CMI-UB}) implies that the r.h.s. of (\ref{new-est}) is an upper bound for the quantities in (\ref{quant}) for all $k$ (regardless of the value of $x_k$). So, the splitting into the sets $N_1$ and $N_2$ is not needed in this case.

The validity of inequality (\ref{omi-fe-cb}) with $\,\varepsilon=\sqrt{\|\Phi-\Psi\|^{E}_{\diamond}}\,$ follows from (\ref{DB-rel+}).

The last assertion of the proposition can be easily proved by noting that $\bar{F}_{H_{A}}(E)=o\shs(\!\sqrt{E})$ as $E\rightarrow+\infty$ by condition (\ref{H-cond+}). $\square$

\begin{corollary}\label{omi-fe-c} \emph{Let $A$ be the $\ell$-mode quantum oscillator with the frequencies $\,\omega_1,...,\omega_{\ell}\,$ and the assumptions of Proposition \ref{omi-fe} hold. Let $F_{\ell,\omega}(E)=\ell\shs[\log((E+E_0)/\ell E_*)+1]$ be a $\varepsilon$-sharp upper bound for $F_{H_A}(E)$, where $E_0\doteq\frac{1}{2}\sum_{i=1}^{\ell}\hbar\omega_i\,$ and $\;E_*\doteq\!\left[\prod_{i=1}^{\ell}\hbar\omega_i\right]^{1/\ell}.\,$ Then
\begin{equation}\label{omi-fe-cb-g}
\hspace{-70pt}\left|I(B^n\!\!:\!D|C)_{\Phi^{\otimes n}\otimes\id_{CD}(\rho)}-I(B^n\!\!:\!D|C)_{\Psi^{\otimes n}\otimes\id_{CD}(\rho)}\right|
\leq n(P_{r}(E,\varepsilon)+g(\varepsilon)+2\varepsilon\log 2),
\end{equation}
for any $\,r\in(0,1]\,$ provided that $\,\varepsilon=\beta_E(\Phi,\Psi)\leq 1/r$,
where
\begin{equation}\label{omi-c-g}
\hspace{-30pt}P_{r}(E,\varepsilon)=2\varepsilon (1+2r)F_{\ell,\omega}(E)+4\ell(2+1/r)\eta(\varepsilon r)+4g(\varepsilon r)+6\varepsilon e^{-\ell},\vspace{-5pt}
\end{equation}
$\,\eta(x)=-x\log x\;$. Continuity bound (\ref{omi-fe-cb-g}) with optimal $\,r$ is tight for large E and any given $\,n$.}
\emph{Continuity bound (\ref{omi-fe-cb-g}) also holds with $\,\varepsilon=\sqrt{\|\Phi-\Psi\|_{\diamond}^{E}}\leq 1/r$.}
\end{corollary}\smallskip

\begin{remark}\label{AT}
The parameter $\,r\,$ in (\ref{omi-fe-cb-g}) can be used to optimize this continuity bound for given $E$ and $\varepsilon$.
By choosing $r\sim 1/F_{\ell,\omega}(E)$ and by noting that $F_{\ell,\omega}(E)=F_{H_A}(E)+o(1)$ for large $E$, we see that the r.h.s. of (\ref{omi-fe-cb-g}) can be made not greater than
$$
2\varepsilon nF_{\ell,\omega}(E)+o\shs(F_{\ell,\omega}(E))=2\varepsilon nF_{H_A}(E)+o\shs(F_{H_A}(E))\quad\textrm{as}\quad E \rightarrow +\infty.
$$
\end{remark}

\emph{Proof.} By Remark \ref{omi-fe-r} the continuity bound (\ref{omi-fe-cb-g}) can be derived from
Proposition \ref{omi-fe} by using the function $\bar{F}_{\ell,\omega}$  defined  in (\ref{F-ub+}) as an upper bound for $\bar{F}_{H_A}$ and the corresponding sequence $\hat{\gamma}(d)$ defined  in (\ref{gamma-h}). Lemma \ref{m-l} below implies that the function $\,E\mapsto \bar{F}_{\ell,\omega}(E)/\sqrt{E}\,$ is non-increasing. So, we may set $\,s=0\,$ in formula (\ref{omi-c}).

Let $\delta\in(0,1]$  and $d_{\delta}$ the minimal natural number such that $\displaystyle{ \bar{E}/\hat{\gamma}(d_{\delta})}\leq\delta^2$.
By using the definition of $d_{\delta}$ and  (\ref{gamma-h}) we obtain
$$
\begin{array}{c}
\displaystyle\log d_{\delta}\leq \log\left(\left[\frac{\bar{E}\delta^{-2}+2E_0}{\ell E_* e^{-1}}\right]^{\ell}+1\right)\leq
 \ell\log\left[\frac{\bar{E}\delta^{-2}+2E_0}{\ell E_* e^{-1}}\right]+\left[\frac{\ell E_* e^{-1}}{\bar{E}\delta^{-2}+2E_0}\right]^{\ell}\\\\\displaystyle\leq\ell\log\left[\frac{E+E_0}{\ell E_*\delta^{2}e^{-1}}\right]+e^{-\ell}= F_{\ell,\omega}(E)-2\ell\log\delta+e^{-\ell},
\end{array}
$$
where the second inequality follows from the inequality $\log (1+x)\leq x$ and the third inequality holds, since
$\delta\leq1$ and $2E_0\geq E_*$. Thus, by setting $\delta=r\varepsilon$ and using monotonicity of the function $g(x)$ it is easy to show that $T_{0,t}(E,\varepsilon)$ is bounded above by the quantity $P_r(E,\varepsilon)$ defined in (\ref{omi-c-g}).

The tightness of continuity bound (\ref{omi-fe-cb-g}) with optimal $\,r$  follows from the tightness of continuity bound (\ref{QC-CB-be}) for the quantum capacity. It can be also shown directly by using the erasure channels $\Phi_{1/2}$ and $\Phi_{1/2-x}$ (see the proof of Theorem 2  in Section 8). $\square$

\begin{lemma}\label{m-l} \emph{The function $\,f(x)=x\log(a/x^2+b)\,$ is increasing on $\,(0,+\infty)\,$ provided that $\,a>0$ and $\,b\geq e/2$.}
\end{lemma}

\emph{Proof.} This assertion is proved by calculation of the derivative $f'(x)$ followed by a simple estimation.

\section{Continuity bounds for the output Holevo quantity}

In analysis of information properties of quantum channels we have to consider the output Holevo quantity of a given channel $\Phi:A\rightarrow B$ corresponding to a discrete or continuous  ensemble $\mu$ of input quantum states, i.e. the quantity
\begin{equation*}
\hspace{-20pt}\chi(\Phi(\mu))=\int H(\Phi(\rho)\shs \|\shs \Phi(\bar{\rho}(\mu))\mu (d\rho )=H(\Phi(\bar{\rho}(\mu
)))-\int H(\Phi(\rho))\mu (d\rho ),  
\end{equation*}%
where the second formula is valid under the condition $H(\Phi(\bar{\rho}(\mu)))<+\infty$.

We will consider the output Holevo quantity $\chi(\Phi(\mu))$ as a function of a pair (channel $\Phi$, input ensemble $\mu$) assuming that
\begin{itemize}
  \item the set of discrete ensembles is equipped with one of the metrics $D_0$, $D_*$ and $D_K$ described in Section 2.3.1;
  \item the set of generalized (continuous)  ensembles is equipped with the Kantorovich metric $D_K$ defined in (\ref{K-D-c});
  \item the set of quantum channels is equipped with the Bures distance $\beta$ defined in (\ref{b-dist+}) in the case $\dim\mathcal{H}_A<+\infty$  and with the energy-constrained Bures distance $\beta_E$ defined in (\ref{ec-b-dist}) in the case $\dim\mathcal{H}_A=+\infty$.
\end{itemize}

\subsection{Finite input dimension}

Speaking about the output Holevo quantity $\chi(\Phi(\mu))$ of a channel $\Phi$ with finite-dimensional input space we restrict attention to discrete ensembles $\mu$, i.e. $\,\mu=\{p_i,\rho_i\}$, for which
$$
\chi(\Phi(\mu))=\chi(\{p_i,\Phi(\rho_i)\})\doteq \sum_{i} p_i H(\Phi(\rho_i)\|\Phi(\bar{\rho})),\quad \bar{\rho}=\sum_{i} p_i\rho_i.
$$

Tight continuity bound for the function $(\Phi, \mu)\mapsto \chi(\Phi(\mu))$  depending  on the input  dimension of a channel is presented in the following proposition.

\begin{property}\label{HQ-CB} \emph{Let $\,\Phi$ and $\,\Psi$ be  quantum channels from a finite-dimensional system $A$ to arbitrary system $B$.
Let $\,\mu$  and  $\,\nu$ be discrete ensembles of states in $\,\mathfrak{S}(\mathcal{H}_A)$. Then
\begin{equation}\label{HQ-CB+}
\left|\chi(\Phi(\mu))-\chi(\Psi(\nu))\right|\leq \varepsilon
\log d_A+\varepsilon
\log 2+2g(\varepsilon),
\end{equation}
where $\,d_A\doteq\dim\mathcal{H}_A$, $\,\varepsilon=D_*(\mu,\nu)+\beta(\Phi,\Psi)$ and $\,g(\varepsilon)=(1+\varepsilon)h_2\!\left(\frac{\varepsilon}{1+\varepsilon}\right)$.}

\emph{If $\,\Phi=\Psi$ then the term $\shs\varepsilon
\log 2$ in (\ref{HQ-CB+}) can be removed. If $\,\mu=\nu$ then (\ref{HQ-CB+}) holds without factor  $2$ in the last term.}

\emph{Continuity bound (\ref{HQ-CB+}) is tight in both cases $\,\Phi=\Psi$ and $\,\mu=\nu$. The metric $D_*$ in (\ref{HQ-CB+}) can be replaced by any of the metrics $D_0$ and $D_K$, the Bures distance $\beta(\Phi,\Psi)$  can be replaced by $\sqrt{\|\Phi-\Psi\|_{\diamond}}$.}
\end{property}

\emph{Proof.}  Assume that $\mu=\{p_i,\rho_i\}$ and $\nu=\{q_i,\sigma_i\}$. Take any
$\epsilon>0$. Let  $\,\{\tilde{p}_i,\tilde{\rho}_i\}$ and $\{\tilde{q}_i,\tilde{\sigma}_i\}$ be ensembles belonging respectively to the sets $\mathcal{E}(\{p_i,\rho_i\})$ and $\mathcal{E}(\{q_i,\sigma_i\})$ such that
\begin{equation}\label{e-1}
D_*(\{p_i,\rho_i\}, \{q_i,\sigma_i\})\geq D_0(\{\tilde{p}_i,\tilde{\rho}_i\}, \{\tilde{q}_i,\tilde{\sigma}_i\})-\epsilon
\end{equation}
(see the definition (\ref{f-metric}) of $D_*$). By Theorem 1 in \cite{Kr&W} there is a common Stinespring representation (\ref{c-S-r}) such that $\|V_{\Phi}-V_{\Psi}\|=\beta(\Phi,\Psi)$. Consider the $qc\shs\textrm{-}$states
$$
\hat{\rho}=\sum_{i} \tilde{p}_iV_{\Phi}\tilde{\rho}_iV_{\Phi}^*\otimes |i\rangle\langle i|\quad
\textrm{and}\quad \hat{\sigma}=\sum_{i}\tilde{q}_iV_{\Psi}\tilde{\sigma}_iV_{\Psi}^*\otimes |i\rangle\langle i|
$$
in $\mathfrak{S}(\mathcal{H}_{BED})$, where $\{|i\rangle\}$ is a basic in $\mathcal{H}_D$. Representation (\ref{chi-rep}) implies that
$$
\chi(\{p_i,\Phi(\rho_i)\})=\chi(\{\tilde{p}_i,\Phi(\tilde{\rho}_i)\})=I(B\!:\!D)_{\hat{\rho}},
$$
$$
\chi(\{q_i,\Psi(\sigma_i)\})=\chi(\{\tilde{q}_i,\Psi(\tilde{\sigma}_i)\})=I(B\!:\!D)_{\hat{\sigma}},
$$
where the first equalities follow from the obvious observation:
$$
\tilde{\mu}\in \mathcal{E}(\mu)\quad\Rightarrow\quad\Theta(\tilde{\mu})\in \mathcal{E}(\Theta(\mu))\quad\Rightarrow\quad \chi(\Theta(\tilde{\mu}))=\chi(\Theta(\mu))
$$
valid for any ensemble $\mu$ and any channel $\Theta$.

Lemma \ref{sl} implies that
\begin{equation}\label{norm+}
\hspace{-45pt} \|\shs\hat{\rho}-\hat{\sigma}\|_1\leq \sum_{i}\|\tilde{p}_i\tilde{\rho}_i-\tilde{q}_i\tilde{\sigma}_i\|_1+2\|V_{\Phi}-V_{\Psi}\|\leq 2(D_{*}(\mu,\nu)+\beta(\Phi,\Psi)+\epsilon).
\end{equation}
Since the states  $\hat{\rho}_{BE}$ and $\hat{\sigma}_{BE}$
are supported on the subspace $V_{\Phi}\mathcal{H}_A\vee V_{\Psi}\mathcal{H}_A$\break of $\mathcal{H}_{BE}$ whose dimension does not exceed $2d_A$,
Lemma \ref{S-CMI-CB}A and (\ref{e-1}),(\ref{norm+}) imply (\ref{HQ-CB+}). If $\Phi=\Psi$ then the above states
$\hat{\rho}_{BE}$ and $\hat{\sigma}_{BE}$ are supported on the $d_A\textrm{-}$dimensional subspace  $V_{\Phi}\mathcal{H}_A=V_{\Psi}\mathcal{H}_A$. If $\,\mu=\nu$ then $\,\hat{\rho}_{D}=\hat{\sigma}_{D}\,$ and hence (\ref{HQ-CB+}) holds with the term $2g(\varepsilon)$ replaced by $g(\varepsilon)$ by Lemma \ref{S-CMI-CB}C.

The tightness of continuity bound (\ref{HQ-CB+}) in the case $\,\Phi=\Psi$ follows from the tightness of the continuity bound (\ref{CHI-CB+}), see Proposition 16 in \cite{CHI}.

The tightness of continuity bound (\ref{HQ-CB+}) in the case $\,\mu=\nu$  follows from the tightness of continuity bound (\ref{HC-CB}) for the Holevo capacity (which is derived from (\ref{HQ-CB+})). It can be also shown directly by using the erasure channels $\Phi_{1/2}$ and $\Phi_{1/2-x}$ (see the proof of Theorem \ref{cap-tcb}  in Section 7).

Since the function $g(x)$ is increasing, the last assertion of the proposition follows from inequalities (\ref{d-ineq}),(\ref{d-ineq+}) and  the right inequality in (\ref{DB-rel}). $\square$

\subsection{Finite input energy}

Speaking about the output Holevo quantity $\chi(\Phi(\mu))$ of a  channel $\Phi$ between infinite-dimensional quantum systems  $A$ and $B$ we will assume that $\mu$ runs over the set $\mathcal{P}(\mathcal{H}_A)$ of all  generalized (continuous) ensembles equipped with the  Kantorovich metric $D_K$ defined in (\ref{K-D-c}) generating the topology of weak convergence on this set. Specifications concerning the case of discrete ensembles will be given as additional remarks.

We will analyse the function $(\Phi,\mu)\mapsto \chi(\Phi(\mu))$ under the constraint on the average energy of $\mu$, i.e. under the condition
$$
E(\mu)\doteq \mathrm{Tr} H_A\bar{\rho}(\mu)=\int\mathrm{Tr} H_A\rho\shs\mu(d\rho)\leq E,
$$
where $H_A$ is the Hamiltonian of the system $A$ and $E\geq E_0\doteq\inf\limits_{\|\varphi\|=1}\langle\varphi|H_A|\varphi\rangle$.

Continuity bound for the function $(\Phi,\mu)\mapsto \chi(\Phi(\mu))$
under the constraint on the average energy of $\,\mu\,$ can be obtained by combining
continuity bounds for the functions $\,\mu\mapsto \chi(\Phi(\mu))\,$ and
$\,\Phi\mapsto \chi(\Phi(\mu))\,$ \emph{not depending}  on $\Phi$ and $\mu$ presented in the following two propositions.

\begin{property}\label{HQ-C-be-1} \emph{Let $\,\Phi:A\rightarrow B$ be an arbitrary quantum channel.
If the Hamiltonian $H_{A}$ of the input system $A$ satisfies condition (\ref{H-cond+}) then the function
$\,\mu\mapsto \chi(\Phi(\mu))$ is uniformly continuous on the subset of $\,\mathcal{P}(\mathcal{H}_A)$ consisting of ensembles $\,\mu\,$
with bounded average energy $\,E(\mu)\doteq \mathrm{Tr} H_A\bar{\rho}(\mu)$. Quantitatively,
\begin{equation}\label{HQ-CB-fe-1}
\left|\chi(\Phi(\mu))-\chi(\Phi(\nu))\right|\leq 2\sqrt{2\varepsilon} \bar{F}_{H_{A}}\!\left(\bar{E}/\varepsilon\right)+2g(\sqrt{2\varepsilon})
\end{equation}
for any ensembles $\shs\mu$ and $\shs\nu$ such that $\,E(\mu),E(\nu)\leq E$ and  $\;D_K(\mu,\nu)\leq\varepsilon\leq 1$, where $\bar{E}=E-E_0$ and $\bar{F}_{H_{A}}$  is the function defined in (\ref{F-bar}).}

\emph{If $\,\mu\shs$ and $\,\nu\shs$ are discrete ensembles then  the Kantorovich metric $D_K$ can be replaced by any of the metrics  $\,D_0$ and $\,D_*$ (described in Section 2.3.1).}

\emph{If $\,A$ is the $\ell$-mode quantum oscillator then the function $\bar{F}_{H_A}$ in (\ref{HQ-CB-fe-1}) can be replaced by its upper bound $\bar{F}_{\ell,\omega}$ defined in (\ref{F-ub+})}.
\end{property}

\emph{Proof.} For discrete ensembles $\mu$ and $\nu$ the inequality (\ref{HQ-CB-fe-1}) with $\varepsilon$ in $[D_*(\mu,\nu),1]$  is proved in \cite[Proposition 7]{AFM}. It follows from (\ref{d-ineq}) and (\ref{d-ineq+}) that this inequality holds for any $\;\varepsilon\in[D(\mu,\nu),1]$, where $D$ is either $D_0$ or $D_K$.

Let $\mu$ and $\nu$ be arbitrary generalized ensembles. The construction from the proof of Lemma 1 in \cite{H-Sh-2} and  the lower semicontinuity of the function $\mu\mapsto\chi(\Phi(\mu))$ (\cite[Proposition 1]{H-Sh-2}) allow to obtain sequences $\{\mu_n\}$ and $\{\nu_n\}$ of discrete ensembles weakly converging, respectively, to $\mu$ and $\nu$ such that
\begin{equation}\label{l-rel}
\lim_{n\rightarrow\infty}\chi(\Phi(\mu_n))=\chi(\Phi(\mu)),\quad\lim_{n\rightarrow\infty}\chi(\Phi(\nu_n))=\chi(\Phi(\nu))
\end{equation}
and $\bar{\rho}(\mu_n)=\bar{\rho}(\mu),\;\bar{\rho}(\nu_n)=\bar{\rho}(\nu)$ for all $n$.  Since inequality (\ref{HQ-CB-fe-1}) holds for the ensembles $\mu_n$ and $\nu_n$ for all $n$ and $D_K(\mu_n,\nu_n)$ tends to $D_K(\mu,\nu)$ as $\,n\rightarrow+\infty$, relations (\ref{l-rel}) imply the validity of (\ref{HQ-CB-fe-1}) for the ensembles $\mu$ and $\nu$. $\square$

\begin{property}\label{HQ-C-be-2} \emph{ Let  $\,\mu$ be an arbitrary  ensemble in $\mathcal{P}(\mathcal{H}_A)$ such that $\,E(\mu)\doteq \mathrm{Tr} H_A\bar{\rho}(\mu)\leq E$.
If the Hamiltonian $H_{A}$ of the input system $A$ satisfies condition (\ref{H-cond+}) then the function $\,\Phi\mapsto\chi(\Phi(\mu))\,$ is  uniformly continuous on the set of all quantum channels from $A$ to any system $B$ with respect to the strong convergence (\ref{sc-def}). Quantitatively,
\begin{equation}\label{HQ-CB-fe-2}
\left|\chi(\Phi(\mu))-\chi(\Psi(\mu))\right|\leq\,T_{s,0}(E,\varepsilon)+g(\varepsilon)+2\varepsilon\log2
\end{equation}
for any channels $\,\Phi$ and $\,\Psi$, where
$\,\varepsilon=\beta_E(\Phi,\Psi)\,$ and $\,T_{s,0}(E,\varepsilon)$ is the quantity defined in Proposition \ref{omi-fe}.} \emph{Inequality (\ref{HQ-CB-fe-2}) also holds with $\,\varepsilon=\sqrt{\|\Phi-\Psi\|_{\diamond}^{E}}$.}

\emph{If $\,A$ is the $\ell$-mode quantum oscillator  then  (\ref{HQ-CB-fe-2}) holds with $\,T_{s,0}(E,\varepsilon)$  replaced by the quantity    $\,P_r(E,\varepsilon)$  defined in (\ref{omi-c-g}) for any $\,r\in(0,1]$ such that $\,\varepsilon\leq 1/r$, where  $\,\varepsilon=\beta_E(\Phi,\Psi)\,$ or $\;\varepsilon=\sqrt{\|\Phi-\Psi\|_{\diamond}^{E}}$.}
\end{property}

\emph{Proof.} Continuity bound  (\ref{HQ-CB-fe-2}) implies the first assertion of the proposition by the last assertions of Propositions \ref{ec-b-d-p} and \ref{omi-fe}.

If $\shs\mu\shs$ is a discrete ensemble then inequality (\ref{HQ-CB-fe-2}) is derived  from Proposition \ref{omi-fe} with $n=1$ and trivial $C$ by using representation (\ref{chi-rep}).

If $\shs\mu\shs$ is an arbitrary ensemble then the validity of (\ref{HQ-CB-fe-2}) can be proved by approximation (by the same way as in the proof of Proposition \ref{HQ-C-be-1}).

The last assertion of the proposition follows from Corollary \ref{omi-fe-c}. $\square$

\section{Continuity bounds for basic capacities of channels with finite input dimension.}

Continuity bounds for basic capacities of quantum channels with finite output dimension $d_B$ are obtained by Leung and Smith in \cite{L&S}. The main term in all these bounds has the form $C\varepsilon\log d_B$, where $C$ is a constant and $\varepsilon$ is a distance between two channels (the  diamond norm of their difference). These continuity bounds are essentially refined in \cite{CHI} by applying the modification of the Leung-Smith approach (consisting in using the quantum conditional mutual information instead of the conditional entropy).

In this section we consider quantum channels with finite input  dimension $d_A$ and  obtain
continuity bounds for basic capacities of such channels with the main term $C\varepsilon\log d_A$, where $\varepsilon$ is the Bures distance between quantum channels described in Section 3.

The \emph{Holevo capacity} of a quantum channel
$\,\Phi:A\rightarrow B\,$  is
defined as
\begin{equation}\label{HC-def}
C_{\chi}(\Phi)=\sup_{\{p_i,\rho_i\}}\chi(\{p_i,\Phi(\rho_i)\}),
\end{equation}
where the supremum is over all discrete ensembles of input states. This quantity characterizes the ultimate rate of classical information transmission   through a channel
provided that nonentangled input encoding is used \cite{H-SCI,Wilde}.

By the Holevo-Schumacher-Westmoreland theorem (cf.\cite{H,S-W}) the \emph{classical capacity} of
a quantum channel $\,\Phi:A\rightarrow B\,$  is given by
the  regularized expression
\begin{equation}\label{CC-def}
C(\Phi)=\lim_{n\rightarrow +\infty }n^{-1}C_{\chi}(\Phi^{\otimes n}).
\end{equation}
If the weak additivity property holds for the channel $\Phi$ then $C(\Phi)=C_{\chi}(\Phi)$ \cite{H-SCI,King,Amosov}.

The \emph{entanglement-assisted classical capacity} of a quantum channel characterizes the ultimate rate of classical information transmission provided
that an entangled state between the input and the output of a channel is used as an
additional resource (see details in \cite{H-SCI,Wilde}). By the Bennett-Shor-Smolin-Thaplyal
theorem (cf.~\cite{BSST}) the 
entanglement-assisted classical capacity of a  channel
$\,\Phi:A\rightarrow B\,$ is given by the expression
\begin{equation}\label{EAC-def}
C_{\mathrm{ea}}(\Phi)=\sup_{\rho \in
\mathfrak{S}(\mathcal{H}_A)}I(\Phi, \rho),
\end{equation}
in which
\begin{equation}\label{mi-def}
 I(\Phi,\rho)=I(B\!:\!R)_{\Phi\otimes\mathrm{Id}_{R}(\hat{\rho})},
\end{equation}
where $\mathcal{H}_R\cong\mathcal{H}_A$ and $\hat{\rho}\shs$
is a pure state in $\mathfrak{S}(\mathcal{H}_{AR})$ such that $\hat{\rho}_{A}=\rho$.
The quantity  $\shs I(\Phi, \rho)\shs$ is called the quantum mutual information of a channel $\Phi$ at a state $\rho$ \cite{H-SCI}.

The \emph{quantum capacity} of a channel characterizes the ultimate rate of quantum information  transmission through a channel (see details in \cite{H-SCI,Wilde}). By the Lloyd-Devetak-Shor  theorem (cf.~\cite{D,L}) the quantum capacity of
a  channel $\,\Phi:A\rightarrow B\,$  is given by
the  regularized expression
\begin{equation}\label{QC-def}
Q(\Phi)=\lim_{n\rightarrow +\infty }n^{-1}\bar{Q}(\Phi^{\otimes n}),
\end{equation}
in which  $\bar{Q}(\Phi)$ denotes the maximal value of the quantum coherent information $I_c(\Phi,\rho)\doteq H(\Phi(\rho))-H(\widehat{\Phi}(\rho))$ over all input states $\rho\in \mathfrak{S}(\mathcal{H}_A)$ (where $\widehat{\Phi}$ is the complementary channel to the channel $\Phi$ defined in (\ref{c-channel})).

The \emph{private capacity} is the capacity of a quantum channel for classical communication with the
additional requirement that almost no information is sent to the environment (see details in \cite{H-SCI,Wilde}). By the Devetak theorem
(cf.~\cite{D}) the private capacity of a  channel $\,\Phi:A\rightarrow B\,$  is given by
the regularized expression
\begin{equation}\label{PCC-def}
C_\mathrm{p}(\Phi)=\lim_{n\rightarrow +\infty }n^{-1}\bar{C}_\mathrm{p}(\Phi^{\otimes n}),
\end{equation}
where
\begin{equation}\label{PHC-def}
\bar{C}_\mathrm{p}(\Phi)=\sup_{\{p_i,\rho_i\}}\left[\chi(\{p_i,\Phi(\rho_i)\})-\chi(\{p_i,\widehat{\Phi}(\rho_i)\})\right]
\end{equation}
(the supremum is over all discrete ensembles of input states and $\,\widehat{\Phi}\,$ is the complementary channel to the channel $\Phi$ defined in (\ref{c-channel})).

Now we consider continuity bounds for all the above capacities depending on the input dimension $d_A$. For the entanglement-assisted classical capacity
the tight continuity bound
\begin{equation*}
\left|\shs C_{\mathrm{ea}}(\Phi)-C_{\mathrm{ea}}(\Psi)\right|\leq 2\varepsilon
\log d_A +g(\varepsilon),
\end{equation*}
where $\,\varepsilon=\frac{1}{2}\|\Phi-\Psi\|_{\diamond}$, is obtained in \cite[Proposition 29]{CHI}. It is easy to show (by using formula (\ref{ci-dif}) below with $\,n=1$) that the same continuity bound holds for the quantity $\,\bar{Q}(\Phi)$ coinciding with $\,Q(\Phi)$ for degradable channels $\Phi$ \cite{DC}.

For others basic capacities tight and close-to-tight
continuity bounds depending on input dimension are presented in the following theorem, in which the Bures distance $\beta(\Phi,\Psi)$
described in Section 3 is used as a measure of divergence between channels $\Phi$ and $\Psi$ and its is assumed that expressions (\ref{HC-def})-(\ref{PHC-def}) remain valid in the case $\dim\mathcal{H}_B=+\infty$.

\begin{theorem}\label{cap-tcb} \emph{Let $\,\Phi$ and $\,\Psi$ be quantum  channels from a finite-dimensional system $A$ to arbitrary system $B$.  Then
\begin{equation}\label{HC-CB}
|\shs C_{\chi}(\Phi)-C_{\chi}(\Psi)|\leq \varepsilon
\log d_A + \varepsilon
\log 2 + g(\varepsilon),\quad\;\;\,
\end{equation}
\begin{equation}\label{CC-CB}
\left|\shs C(\Phi)-C(\Psi)\right|\leq 2\varepsilon
\log d_A +2\varepsilon
\log 2+g(\varepsilon),
\end{equation}
\begin{equation}\label{QC-CB}
\left|\shs Q(\Phi)-Q(\Psi)\right|\leq 2\varepsilon
\log d_A +2\varepsilon
\log 2+g(\varepsilon),
\end{equation}
\begin{equation}\label{PHC-CB}
|\shs\bar{C}_{\mathrm{p}}(\Phi)-\bar{C}_{\mathrm{p}}(\Psi)|\leq 2\varepsilon
\log d_A+2\varepsilon
\log 2 +2g(\varepsilon),\;
\end{equation}
\begin{equation}\label{PCC-CB}
\left|\shs C_{\mathrm{p}}(\Phi)-C_{\mathrm{p}}(\Psi)\right|\leq 4\varepsilon
\log d_A +4\varepsilon
\log 2+2g(\varepsilon),\;
\end{equation}
where  $d_A\doteq\dim\mathcal{H}_A$, $\,\varepsilon=\beta(\Phi,\Psi)\,$  and $\;g(\varepsilon)=(1+\varepsilon)h_2\!\left(\frac{\varepsilon}{1+\varepsilon}\right)$.}

\emph{The continuity bounds (\ref{HC-CB}),(\ref{QC-CB}) and (\ref{PHC-CB}) are tight, other continuity bounds  are close-to-tight (up to factor 2 in the main term). The Bures  distance $\beta(\Phi,\Psi)$ in  (\ref{HC-CB})-(\ref{PCC-CB})  can be replaced by  $\sqrt{\|\Phi-\Psi\|_{\diamond}}$.}
\end{theorem}\smallskip

\emph{Proof.} Continuity bound (\ref{HC-CB}) directly follows from Proposition \ref{HQ-CB} in Section 6.1 and the definition of the Holevo capacity.

Continuity bound (\ref{CC-CB}) is  obtained by using  representation (\ref{chi-rep}), Proposition \ref{omi} in Section 5.2 and Lemma 12 in \cite{L&S}.

Continuity bound (\ref{QC-CB}) is proved by representing the coherent information  as follows
$\,I_c(\Phi,\rho)= I(B\!:\!R)_{\Phi\otimes\id_R}(\hat{\rho})-H(\rho)$,
where $\hat{\rho}$ is a purification in $\mathfrak{S}(\mathcal{H}_{AR})$ of a state $\rho$. For arbitrary quantum channels $\Phi$ and $\Psi$, any $n$ and a state $\rho$ in $\mathfrak{S}(\mathcal{H}_{A^n})$ this  implies
\begin{equation}\label{ci-dif}
\hspace{-30pt}I_c(\Phi^{\otimes n}\!\!,\rho)-I_c(\Psi^{\otimes n}\!\!,\rho)=I(B^n\!:\!R)_{\Phi^{\otimes n}\otimes\id_{R}}(\hat{\rho})-I(B^n\!:\!R)_{\Psi^{\otimes n}\otimes\id_{R}}(\hat{\rho}),
\end{equation}
where $\hat{\rho}$ is a purification  in $\mathfrak{S}(\mathcal{H}_{A^nR})$ of the state $\rho$. This expression, Proposition \ref{omi} in Section 5.2 and Lemma 12 in \cite{L&S} imply (\ref{QC-CB}).

Continuity bound (\ref{PHC-CB}) is  obtained by using  Proposition \ref{HQ-CB} in Section 6.1 twice and  Corollary \ref{ec-b-d-c2}A in Section 3.

Continuity bound (\ref{PCC-CB}) is proved by noting that representation (\ref{chi-rep}) implies
\begin{equation}\label{PHC-def+}
\bar{C}_\mathrm{p}(\Phi^{\otimes n})=\sup_{\hat{\rho}} \left[I(B^n\!:\!C)_{\Phi^{\otimes n}\otimes\id_{C}(\hat{\rho})}-I(E^n\!:\!C)_{\widehat{\Phi}^{\otimes n}\otimes\id_{C}(\hat{\rho})}\right],
\end{equation}
where the supremum is over all $qc$-states in $A^nC$. Inequality (\ref{PCC-CB}) is obtained by using Corollary \ref{ec-b-d-c2}A in Section 3, Proposition \ref{omi} in Section 5.2 twice and Lemma 12 in \cite{L&S}.

To prove the tightness of continuity bounds (\ref{HC-CB}),(\ref{QC-CB}) and (\ref{PHC-CB}) consider the family of erasure channels
\begin{equation}\label{er-ch}
\Phi_p(\rho)=\left[\begin{array}{cc}
(1-p)\rho &  0 \\
0 &  p\mathrm{Tr}\rho
\end{array}\right], \quad p\in[0,1].
\end{equation}
from $d$-dimensional quantum system $A$ to $(d+1)$-dimensional quantum system $B$. It is well known (see \cite{H-SCI,Wilde}) that
\begin{equation}\label{ecc}
  C(\Phi_p)=C_{\chi}(\Phi_p)=(1-p)\log d,
\end{equation}
\begin{equation}\label{ecc+}
  Q(\Phi_p)=C_\mathrm{p}(\Phi_p)=\bar{C}_\mathrm{p}(\Phi_p)=\max\{(1-2p)\log d, 0\}.
\end{equation}
By writing the  channel $\Phi_p$ as the map $\rho\mapsto (1-p)\rho\oplus [p\mathrm{Tr}\rho]|\psi\rangle\langle\psi|$ from $\mathfrak{T}(\mathcal{H}_A)$ to $\mathfrak{T}(\mathcal{H}_A\oplus\mathcal{H}_{\psi})$, where $\mathcal{H}_{\psi}$ is the space generated by $|\psi\rangle$, we see that the isometry
$$
V_p:|\varphi\rangle\mapsto\sqrt{1-p}|\varphi\rangle\otimes|\psi\rangle\oplus\sqrt{p}|\psi\rangle\otimes|\varphi\rangle
$$
from $\mathcal{H}_A$ into $\mathcal{H}_{BE}$, where $\mathcal{H}_{E}=\mathcal{H}_{B}=\mathcal{H}_A\oplus\mathcal{H}_{\psi}$, is a Stinespring isometry for $\Phi_p$, i.e.
$\Phi_p(\rho)=\mathrm{Tr}_E V_{p}\rho V^*_{p}$, for each $p$. Direct calculation shows that
\begin{equation}\label{dV}
\hspace{-30pt} \|V_{1/2-x}-V_{1/2}\|=\sqrt{2-\sqrt{1-2x}-\sqrt{1+2x}}=x+o(x)\quad \textrm{as }\;x\rightarrow0.
\end{equation}
It follows from (\ref{ecc}) and (\ref{ecc+}) that $\;C_{*}(\Phi_{1/2-x})-C_{*}(\Phi_{1/2})=x\log d\,$ for $C_*=C_{\chi},C$ and that
$C_*(\Phi_{1/2-x})-C_*(\Phi_{1/2})=2x\log d\,$ for  $C_*=Q,\bar{C}_\mathrm{p},C_\mathrm{p}$.

Since (\ref{dV}) implies $\beta(\Phi_{1/2-x},\Phi_{1/2})\leq x+o(x)$ for small $x$, we see that continuity bounds (\ref{HC-CB}),(\ref{QC-CB}) and (\ref{PHC-CB}) are tight (for large $d_A$), while others are close-to-tight (up to factor 2 in the main term).

The last assertion of the theorem  follows from the right inequality in (\ref{DB-rel}) and monotonicity of the function $g(x)$. $\square$

\section{Continuity bounds for basic capacities of energy-constrained infinite-dimensional channels.}

In \cite{SCT,W-EBN} continuity bounds for classical and quantum capacities  of energy-constrained
infinite-dimensional channels with bounded energy amplification factor are obtained. In this section we use the results of Sections 5 and 6 to
derive tight and close-to-tight continuity bounds for basic capacities valid for \emph{arbitrary} energy-constrained
infinite-dimensional channels.

When we consider transmission of  information over
infinite-dimensional quantum channels we have to impose constraints on states
used for encoding information to be consistent with the physical
implementation of the process. A typical physically motivated constraint is
the boundedness of the average energy of states used for encoding information.
For a single channel from $A$ to $B$ this constraint is expressed by the inequality
\begin{equation}\label{lc}
\mathrm{Tr}H_A\rho \leq E,\quad \rho\in\mathfrak{S}(\mathcal{H}_A),
\end{equation}%
where $H_A$ is  the Hamiltonian of the
input system $A$, for $n$-copies of a channel it
can be written as follows
\begin{equation}\label{lc+}
\mathrm{Tr}H_{A^n}\rho \leq nE,\quad \rho\in\mathfrak{S}(\mathcal{H}^{\otimes n}_{A}),
\end{equation}%
where $\,H_{A^n}=H_A\otimes I_A\otimes\ldots\otimes I_A+\ldots+I_A\otimes \ldots\otimes I_A\otimes H_A\, $ is the Hamiltonian of the system $A^n$
 \cite{H-SCI,H-c-w-c,Wilde+}.

In this section we apply the results of Sections 5 and 6 to obtain continuity bounds for capacities of infinite-dimensional quantum channels with constraint (\ref{lc+}) depending on $E$ and the energy-constrained Bures distance between channels (introduced in Section 3) assuming that the Hamiltonian $H_A$ satisfies condition (\ref{H-cond+}). We consider all the basic capacities excepting the classical
entanglement-assisted capacity $C_{\mathrm{ea}}$, since the continuity bound for this capacity depending on the input energy bound and the energy-constrained diamond norm distance between quantum channels is obtained in \cite[Proposition 7]{SCT}.

The Holevo capacity of any channel
$\,\Phi:A\rightarrow B\,$ with constraint (\ref{lc+}) is defined as:
\begin{equation}\label{chi-cap-def-be}
C_{\chi}(\Phi,H_A,E)=\sup_{\mathrm{Tr} H_A\bar{\rho}\leq E}\chi(\{p_i,\Phi(\rho_i)\}),
\end{equation}%
where the supremum is over all discrete ensembles of states in $\mathfrak{S}(\mathcal{H}_{A})$ with the average energy not exceeding $E$.

By the Holevo-Schumacher-Westmoreland theorem adapted for constrained infinite-dimensional channels (see \cite[Proposition 3]{H-c-w-c}) the classical capacity of
any  channel $\,\Phi:A\rightarrow B\,$  with constraint (\ref{lc+}) is given by
the  regularized expression
\begin{equation*}
C(\Phi,H_A,E)=\lim_{n\rightarrow +\infty }n^{-1}C_{\chi}(\Phi^{\otimes n},H_{A^n},nE).
\end{equation*}

Detailed analysis of the energy-constrained quantum and private capacities in the context of general-type infinite-dimensional channels\footnote{There are many papers devoted to analysis of these capacities for Gaussian channels, see \cite{H&W,Wolf} and the surveys in \cite{Wilde+,W&C}.} has been made recently by Wilde and Qi in \cite{Wilde+}. The results in \cite{Wilde+} give considerable reasons to conjecture validity of the following generalizations of
the Lloyd-Devetak-Shor theorem and of the Devetak theorem to energy-constrained infinite-dimensional channels:
\begin{itemize}
  \item
the quantum capacity of any channel
$\,\Phi:A\rightarrow B\,$ with constraint (\ref{lc+})
is given by
the  regularized expression
\begin{equation*}
Q(\Phi,H_A,E)=\lim_{n\rightarrow +\infty }n^{-1}\bar{Q}(\Phi^{\otimes n},H_{A^n},nE),
\end{equation*}
where $\,\bar{Q}(\Phi,H_A,E)\,$ is the least upper bound of the coherent information $\,I_c(\Phi,\rho)\doteq I(\Phi,\rho)-H(\rho)\,$ on the set of all input states $\rho\in \mathfrak{S}(\mathcal{H}_A)$ satisfying (\ref{lc}).
  \item the private  capacity of any  channel
$\,\Phi:A\rightarrow B\,$ with  constraint (\ref{lc+})
is given by
the  regularized expression
\begin{equation*}
C_\mathrm{p}(\Phi,H_A,E)=\lim_{n\rightarrow +\infty }n^{-1}\bar{C}_\mathrm{p}(\Phi^{\otimes n},H_{A^n},nE),
\end{equation*}
where
$\bar{C}_\mathrm{p}(\Phi,H_A,E)=\sup\limits_{\mathrm{Tr} H_A\bar{\rho}\leq E}\left[\chi(\{p_i,\Phi(\rho_i)\})-\chi(\{p_i,\widehat{\Phi}(\rho_i)\})\right]$
(the supremum is over all discrete ensembles of states in $\mathfrak{S}(\mathcal{H}_{A})$ with the average energy not exceeding $E$ and $\widehat{\Phi}$ is the complementary channel to the channel $\Phi$ defined in (\ref{c-channel})).
\end{itemize}

The uniform finite-dimensional approximation theorem for all the above capacities obtained in \cite{UFA} and the analog of Theorem \ref{cap-tcb} for constrained channels with finite-dimensional input space show that
all the functions
$$
\Phi\mapsto C_*(\Phi,H_A,E),\quad C_*=C_{\chi},\; C,\;\bar{Q},\; Q,\;\bar{C}_\mathrm{p},\;C_\mathrm{p},
$$
are uniformly continuous on the set of all channels from $A$ to arbitrary system $B$ with respect to the strong (pointwise) convergence (\ref{sc-def}) provided that the Hamiltonian $H_A$  satisfies condition (\ref{H-cond+}). But the continuity bounds obtained in \cite[Theorem 2]{UFA} by this way are too rough and can not be used in applications. The following theorem presents more sharp and usable continuity bounds for all the above functions  excepting $\,\bar{Q}(\Phi,H_A,E)\,$ which depend only on the input energy bound $E$ and the energy-constrained Bures distance $\beta_E$ between quantum channels defined in (\ref{ec-b-dist}) (the tight continuity bound for the quantity $\,\bar{Q}(\Phi,H_A,E)\,$ depending on $E$ coincides with the continuity bound for the capacity $\,C_{\mathrm{ea}}(\Phi,H_A,E)\,$ obtained in \cite[Proposition 7]{SCT}).

\begin{theorem}\label{cap-cb-fe} \emph{Let $H_A$ be the Hamiltonian of a system $A$ satisfying condition (\ref{H-cond+}).
Let $\,s=0\,$ if the function $\,E\mapsto \bar{F}_{H_A}(E)/\sqrt{E}\,$ is non-increasing and  $\,s=1$ otherwise.\footnote{The function $\bar{F}_{H_A}$ is defined in (\ref{F-bar}).}  Then
for arbitrary quantum channels $\,\Phi$ and $\,\Psi$ from the system $A$ to any system $B$ the following inequalities hold
\begin{equation}\label{HC-CB-be}
|\shs C_{\chi}(\Phi,H_A,E)-C_{\chi}(\Psi,H_A,E)|\leq T_{s,0}(E,\varepsilon)+g(\varepsilon)+2\varepsilon\log 2,\;\;\;\,
\end{equation}
\begin{equation}\label{CC-CB-be}
\left|\shs C(\Phi,H_A,E)-C(\Psi,H_A,E)\right|\leq T_{s,0}(E,\varepsilon)+g(\varepsilon)+2\varepsilon\log 2,\;
\end{equation}
\begin{equation}\label{QC-CB-be}
\left|\shs Q(\Phi,H_A,E)-Q(\Psi,H_A,E)\right|\leq T_{s,1}(E,\varepsilon)+g(\varepsilon)+2\varepsilon\log 2,
\end{equation}
\begin{equation}\label{PHC-CB-be}
|\shs \bar{C}_\mathrm{p}(\Phi,H_A,E)-\bar{C}_\mathrm{p}(\Psi,H_A,E)|\leq 2T_{s,0}(E,\varepsilon)+2g(\varepsilon)+4\varepsilon\log 2,
\end{equation}
\begin{equation}\label{PCC-CB-be}
\left|\shs C_{\mathrm{p}}(\Phi,H_A,E)-C_{\mathrm{p}}(\Psi,H_A,E)\right|\leq 2T_{s,1}(E,\varepsilon)+2g(\varepsilon)+4\varepsilon\log 2,
\end{equation}
where $\,\varepsilon=\beta_E(\Phi,\Psi)$ and $T_{s,t}(E,\varepsilon)$ is the quantity defined in Proposition \ref{omi-fe}.} \emph{Inequalities  (\ref{HC-CB-be})-(\ref{PCC-CB-be}) also hold with $\,\varepsilon=\sqrt{\|\Phi-\Psi\|^{E}_{\diamond}}$.}

\emph{If $\,A$ is the $\ell$-mode quantum oscillator  then  inequalities  (\ref{HC-CB-be})-(\ref{PCC-CB-be}) hold with $T_{s,t}(E,\varepsilon)$  replaced by the quantity   $\,P_r(E,\varepsilon)$  defined in (\ref{omi-c-g}) for any $r\in(0,1]$ such that $\,\varepsilon\leq 1/r$, where  $\,\varepsilon=\beta_E(\Phi,\Psi)\,$ or $\;\varepsilon=\sqrt{\|\Phi-\Psi\|_{\diamond}^{E}}$. In this case continuity bounds (\ref{QC-CB-be}) with optimal $\,r$ is tight
for large $E$, other continuity bounds are close-to-tight (up to factor 2 in the main term).}
 \end{theorem}

\emph{Proof.} The continuity bound (\ref{HC-CB-be}) for the Holevo capacity  directly follows from its definition (\ref{chi-cap-def-be}) and Proposition \ref{HQ-C-be-2} in Section 6.2.

The continuity bound (\ref{CC-CB-be}) is proved by using  representation (\ref{chi-rep}), Proposition \ref{omi-fe} in Section 5.2 and  Lemma 12 in \cite{L&S}.
The possibility to set $\,t=0\,$ in (\ref{CC-CB-be}) is due to the fact the supremum in the definition of $C_{\chi}(\Phi^{\otimes n},H_{A^n},nE)$ can be taken over all ensembles  $\{p_i,\rho_i\}$ of states in $\mathfrak{S}(\mathcal{H}^{\otimes n}_A)$ with the average state $\bar{\rho}$ such that $\,\mathrm{Tr} H_A\bar{\rho}_{A_k}\leq E\,$ for all $k=\overline{1,n}$. This can be easily shown by using the symmetry arguments (see the proof of Proposition 6 in \cite{SCT}).

The continuity bound (\ref{QC-CB-be}) is proved by using  expression (\ref{ci-dif}), Proposition \ref{omi-fe} in Section 5.2 and  Lemma 12 in \cite{L&S}.

The continuity bound (\ref{PHC-CB-be}) is  obtained by using  Proposition \ref{HQ-C-be-2} in Section 6.2 twice and Corollary \ref{ec-b-d-c2}B in Section 3.

The continuity bound (\ref{PCC-CB-be}) is proved by using representation (\ref{chi-rep}),  Corollary \ref{ec-b-d-c2}B in Section 3, Proposition \ref{omi-fe} in Section 5.2 twice and Lemma 12 in \cite{L&S}.

The assertion concerning the $\ell$-mode quantum oscillator  follows from Corollary  \ref{omi-fe-c} in Section 5.2.
To show that the continuity bounds are tight or close-to-tight consider the family of erasure channels
(\ref{er-ch}) from the $\ell$-mode quantum oscillator to the system described by the space $\mathcal{H}_B=\mathcal{H}_A\oplus\{c\psi\}$.
By generalizing the arguments from \cite[Ch.10]{H-SCI} it is easy to show that $\,C(\Phi_p, H_A, E)=C_{\chi}(\Phi_p, H_A, E)=(1-p)M\,$ and
\begin{equation*}
\hspace{-30pt}  Q(\Phi_p, H_A, E)=C_\mathrm{p}(\Phi_p, H_A, E)=\bar{C}_\mathrm{p}(\Phi_p, H_A, E)=\max\{(1-2p)M, 0\},
\end{equation*}
where $M=F_{H_A}(E)$. It follows  that $\;C_{*}(\Phi_{1/2-x},H_A, E)-C_{*}(\Phi_{1/2},H_A, E)=xM$, $C_*=C_{\chi},C\,$ and that
$\,C_*(\Phi_{1/2-x},H_A,E)-C_*(\Phi_{1/2},H_A, E)=2xM$, $C_*=Q, \bar{C}_\mathrm{p},C_\mathrm{p}$.

By  the proof of Theorem 1  $\beta_E(\Phi_{1/2-x},\Phi_{1/2})\leq\beta(\Phi_{1/2-x},\Phi_{1/2})\leq x+o(x)$ for small $x$. So, by using Remark \ref{AT} in Section 5.2 it is easy to show that  continuity bound (\ref{QC-CB-be}) is tight for large $E$ and others are close-to-tight (up to factor 2 in the main term). $\square$

\ack I am grateful to A.S.Holevo, A.V.Bulinski and G.G.Amosov for useful comments. I am also grateful to the  participants of the workshop "Recent advances in continuous variable quantum information theory", Barcelona, April, 2016  (especially to A.Winter) for stimulating discussion.
Special thanks to M.M.Wilde for valuable communication concerning  capacities of infinite-dimensional channels with energy constraints.

\appendix

\setcounter{section}{1}

\section*{Appendix: the addition to the proof of Proposition \ref{ec-b-d-p}}

 Denote by $\beta'_E(\Phi,\Psi)$ the r.h.s. of (\ref{ec-b-dist+}).
Let $\mathfrak{C}_{H_A,E}$ be the subset of $\mathfrak{S}(\mathcal{H}_A)$ determined by the inequality $\,\mathrm{Tr} H_A\rho\leq E\,$ and $\mathcal{N}(\Phi,\Psi)=\bigcup V_{\Phi}^*V_{\Psi}$, where the union is over \emph{all} common Stinespring representations (\ref{c-S-r}). Then it is easy to see that
\begin{equation}\label{beta-e}
\beta'_E(\Phi,\Psi)=\inf_{N\in\mathcal{N}(\Phi,\Psi)} \sup_{\rho\in\mathfrak{C}_{H_A,E}}\sqrt{2-2\Re\shs\mathrm{Tr} N\rho}
\end{equation}
Following the proof of Theorem 1 in \cite{Kr&W} show that $\mathcal{N}(\Phi,\Psi)$ coincides with the set
$$
\mathcal{M}(\Phi,\Psi)\doteq \{V_{\Phi} ^*(I_B\otimes C)V_{\Psi}\,|\,C\in \mathfrak{B}(\mathcal{H}_E), \|C\|\leq 1\},
$$ defined via some \emph{fixed} common Stinespring representation (\ref{c-S-r}). It would imply, in particular, that
$\mathcal{M}(\Phi,\Psi)$ does not depend on this representation.

To show that $\mathcal{M}(\Phi,\Psi)\subseteq\mathcal{N}(\Phi,\Psi)$ it suffices to find for any contraction $C\in\mathfrak{B}(\mathcal{H}_E)$ a common Stinespring representation for  $\Phi$ and $\Psi$ with the isometries
$\tilde{V}_{\Phi}$ and $\tilde{V}_{\Psi}$ from $\mathcal{H}_A$ to $\mathcal{H}_B\otimes\mathcal{H}_{\tilde{E}}$ such that $\tilde{V}^*_{\Phi}\tilde{V}_{\Psi}=V_{\Phi} ^*(I_B\otimes C)V_{\Psi}$.

Let $\mathcal{H}_{\tilde{E}}=\mathcal{H}^1_E\oplus\mathcal{H}^2_E$, where $\mathcal{H}^1_E$ and $\mathcal{H}^2_E$ are copies of $\mathcal{H}_E$. For given $C$ define the isometries
$\tilde{V}_{\Phi}$ and $\tilde{V}_{\Psi}$ from $\mathcal{H}_A$ into $\mathcal{H}_B\otimes(\mathcal{H}_{E_1}\oplus\mathcal{H}_{E_2})$ by setting
$$
\tilde{V}_{\Phi}|\varphi\rangle=V_{\Phi}|\varphi\rangle\oplus|0\rangle,\quad
\tilde{V}_{\Psi}|\varphi\rangle=(I_B\otimes C)V_{\Psi}|\varphi\rangle\oplus \left(I_B\otimes\sqrt{I_{E}-C^*C}\right)V_{\Psi}|\varphi\rangle
$$
for any $\varphi\in\mathcal{H}_A$,
where we assume that the isometries $V_{\Phi}$ and $V_{\Psi}$ act from $\mathcal{H}_A$ to $\mathcal{H}_B\otimes \mathcal{H}^1_E$ and
$\mathcal{H}_B\otimes \mathcal{H}^2_E$ correspondingly, while the contraction $C$ acts from $\mathcal{H}^2_E$ to $\mathcal{H}^1_E$.  It is easy to see that
the isometries $\tilde{V}_{\Phi}$ and $\tilde{V}_{\Psi}$ form a common Stinespring representation for the channels $\Phi$ and $\Psi$ with the required property.

To prove that $\mathcal{N}(\Phi,\Psi)\subseteq\mathcal{M}(\Phi,\Psi)$ take any common Stinespring representation for  $\Phi$ and $\Psi$ with the isometries
$\tilde{V}_{\Phi}$ and $\tilde{V}_{\Psi}$ from $\mathcal{H}_A$ to $\mathcal{H}_B\otimes\mathcal{H}_{\tilde{E}}$. By Theorem 6.2.2 in \cite{H-SCI} there exist partial isometries $W_{\Phi}$ and $W_{\Psi}$ from $\mathcal{H}_E$ to $\mathcal{H}_{\tilde{E}}$ such that $\tilde{V}_{\Phi}=(I_B\otimes W_{\Phi})V_{\Phi}$ and $\tilde{V}_{\Psi}=(I_B\otimes W_{\Psi})V_{\Psi}$. So, $\tilde{V}^*_{\Phi}\tilde{V}_{\Psi}=V^*_{\Phi}(I_B\otimes W^*_{\Phi}W_{\Psi})V_{\Psi}\in \mathcal{M}(\Phi,\Psi)$, since $\|W^*_{\Phi}W_{\Psi}\|\leq 1$.

Since $\mathcal{N}(\Phi,\Psi)=\mathcal{M}(\Phi,\Psi)$, the infimum in (\ref{beta-e}) can be taken over the set $\mathcal{M}(\Phi,\Psi)$. This implies
\begin{equation}\label{beta-d}
\hspace{-30pt}\begin{array}{rl}
\beta'_E(\Phi,\Psi)\!&=\,\displaystyle\inf_{C\in\mathfrak{B}_1(\mathcal{H}_E)}\sup_{\rho\in\mathfrak{C}(\mathcal{H}_{A},E)}\sqrt{2-2\Re\shs\mathrm{Tr} V_{\Phi}^*(I_B\otimes C)V_{\Psi}\rho}\\
& =\displaystyle\sup_{\rho\in\mathfrak{C}(\mathcal{H}_{A},E)}\inf_{C\in\mathfrak{B}_1(\mathcal{H}_E)} \sqrt{2-2\Re\shs\mathrm{Tr} V_{\Phi} ^*(I_B\otimes C)V_{\Psi}\rho},
\end{array}
\end{equation}
where the possibility to change the
order of the optimization follows from Ky Fan's minimax theorem \cite{Simons} and the $\sigma$-weak compactness of the unit ball $\mathfrak{B}_1(\mathcal{H}_E)$ of $\mathfrak{B}(\mathcal{H}_E)$ \cite{B&R}. It is easy to see that
\begin{equation}\label{a-eq}
\hspace{-45pt} \sup_{C\in\mathfrak{B}_1(\mathcal{H}_E)}\Re\,\mathrm{Tr} V_{\Phi}^*(I_B\otimes C)V_{\Psi}\rho  = \sup_{C\in\mathfrak{B}_1(\mathcal{H}_E)}|\langle V_{\Phi}\otimes I_R\shs\varphi|I_{BR}\otimes C |V_{\Psi}\otimes I_R \shs\varphi\rangle|,
\end{equation}
where $R$ is any system and $\varphi$ is a purification of $\rho$, i.e. a  vector in $\mathcal{H}_{AR}$ such that $\mathrm{Tr}_R|\varphi\rangle\langle\varphi|=\rho$.

Since for \emph{any} common Stinespring representation (\ref{c-S-r}) and any system $R$ the vectors $V_{\Phi}\otimes I_R\shs|\varphi\rangle$ and $V_{\Psi}\otimes I_R\shs|\varphi\rangle$ in $\mathcal{H}_{BER}$ are purifications of the states
$\Phi\otimes \id_R(|\varphi\rangle\langle\varphi|)$ and $\Psi\otimes \id_R(|\varphi\rangle\langle\varphi|)$ in $\mathfrak{S}(\mathcal{H}_{BR})$, by using the relation $\mathcal{N}(\Phi,\Psi)=\mathcal{M}(\Phi,\Psi)$ proved before and Uhlmann's theorem \cite{Uhlmann}\cite[Ch.9]{Wilde} it is easy to show that the quantity in the r.h.s. of (\ref{a-eq}) coincides with the fidelity of the states $\Phi\otimes \id_R(|\varphi\rangle\langle\varphi|)$ and $\Psi\otimes \id_R(|\varphi\rangle\langle\varphi|)$. This and (\ref{beta-d}) implies that $\beta'_E(\Phi,\Psi)=\beta_E(\Phi,\Psi)$.

Assertion c) can be derived from the attainability of the infima in (\ref{beta-d}) which follows from the $\sigma$-weak compactness of the unit ball $\mathfrak{B}_1(\mathcal{H}_E)$.

\end{document}